# Title
Microbial and Viral Ecology Analysis for Metagenomic Data

# Authors
James C. Kosmopoulos[1,2] and Karthik Anantharaman[1,3,*]

[1] Department of Bacteriology, University of Wisconsin-Madison, Madison, Wisconsin, USA
[2] Microbiology Doctoral Training Program, University of Wisconsin-Madison, Madison, Wisconsin, USA
[3] Department of Integrative Biology, University of Wisconsin-Madison, Madison, Wisconsin, USA
[*] Correspondence: karthik@bact.wisc.edu

## Abstract

The explosion in known microbial diversity in the last two decades has made it abundantly clear that microbes in the environment do not exist in isolation, they are members of communities. Accordingly, omics approaches such as metagenomics have revealed that interactions between diverse groups of community members such as archaea, bacteria, and viruses (bacteriophage) are common and have significant impacts on entire microbiomes. Thus, to have a well-developed understanding of microbes as they naturally exist in the environment, biological entities off all kinds must be studied together. While numerous protocols for metagenome analysis exist, comprehensive published protocols for the simultaneous analysis of viruses and prokaryotes together are scarce. Further, as bioinformatic methods for microbiology rapidly advance, existing metagenomic tools and pipelines require frequent reevaluation. This ensures the adherence of best practices for microbiome and metagenomic data analysis. Here, we offer an expansive approach for the joint analysis of bulk sequence data from a mixed microbial community (metagenomes) and viral-sized fraction communities (viromes). This chapter serves as a beginners-level guide for researchers with limited bioinformatics expertise who wish to engage in multi-scale metagenome and virome analyses. We cover steps from initial study design to sequence read processing, metagenome assembly, quality control, virus identification, microbial and viral genome binning, taxonomic characterization, species-level clustering, and host-virus predictions. We also provide the bioinformatic scripts used in our workflow for reuse in one's own computational methods. Lastly, we discuss additional approaches a researcher can take after processing data with this workflow.

## 1 - Introduction

In the current age of genomics, microbiology as a field is rapidly growing and evolving as DNA sequence data becomes more accessible and higher in quality. Accordingly, metagenomics is a powerful tool to study the genetic material of entire microbiomes directly from biological samples. Since these approaches can be cultivation-independent, the application of metagenomics and bioinformatics to the study of microbiomes enables a deeper understanding of microbial ecology, evolution, and the roles that microorganisms play in the environment and human health. While metagenomes of a mixed-community of bacteria, archaea, and viruses offer a broad overview of the biological entities within an environment (1–4), focusing on virus-like-particle sized fraction metagenomes (viromes) can provide an additional layer of focus. Viruses (particularly bacteriophage) play crucial roles in regulating microbial populations (5–7), horizontal gene transfer (8, 9), and influencing microbial genetic diversity (10–12). Viromes can uncover the diversity, abundance, and functions of viruses in microbiomes (4, 13, 14). Therefore, analyzing viromes in conjunction with mixed-community metagenomes can greatly enhance a researcher's understanding of microbial and viral dynamics, interactions, and impacts on their hosts and environments.

To streamline the complex processes involved in metagenomic and virome analyses, several workflows and pipelines have been developed. Tools like MetaWRAP (15), ViWrap (16), the JGI Metagenome Workflow (17), and Anvi'o (18) offer comprehensive suites for processing, analyzing, and visualizing metagenomic data. These tools facilitate quality control, assembly, binning, annotation, and comparative genomics, allowing researchers to derive meaningful insights from their data efficiently. Despite the existence of these useful automated pipelines, a step-by-step manual that is presented in this chapter remains crucial. This manual is intended to serve as a resource for understanding the intricacies of each analytical step in common metagenomics and viromics workflows. We also explain the rationale behind methodological choices and discuss why they may or may not apply to a given research context.

## 2 - Materials

### *2.1 - Hardware and Operating System*

When working with one or two individual samples, performing *in silico* analyses of total microbial and viral communities can typically be done on a personal computer with at least 6 GB of RAM, a processor with at least 4 cores, and 100 GB of available storage. However, the incredible diversity of microbes and their viruses often requires biological replicates across temporal, spatial, and/or physiochemical scales to perform in-depth investigations (see Methods). Scaling bioinformatic analyses with multiple samples will often require the use of a high-performance computing server. These servers are sometimes available to researchers at individual research institutions or access to one can be purchased from an online host. There are also webservers, such as Galaxy (19), that allow users to access resources for bioinformatics for free.

Moreover, the operating system of the machine performing bioinformatic analyses is a critical component. Many bioinformatic tools require a command-line environment (a *shell*) to operate and often have specific OS requirements. The vast majority of tools are compatible with Linux operating systems such as Ubuntu or Debian, and, to a slightly lesser extent Mac OS (Unix). Except for a handful of tools that exist within a graphical user interface or online, most tools will require either a Linux or Mac (Unix) operating system. Researchers running analyses locally (without a server) on a Windows machine can install a virtual machine for free that will allow them to access a Linux operating system, such as Windows Subsystem for Linux (learn.microsoft.com/en-us/windows/wsl).

The analyses performed here were conducted on a server with an Ubuntu 20.04.06 OS with access to 256 CPUs, 2 TB of RAM, and 180 TB of storage. However, the analyses to be run will be capped to use at most 16 CPUs, 100 GB of RAM, and 2 TB of disk space.

### *2.2 - Software*

The software used in the analyses in this chapter are listed in Table 1. The versions of the packages used here are also provided for reproducibility, but installing the same versions on these packages may not be required for all aspects of these analyses.

**Table 1. List of tools used in Section 3.**

| Tool | Version | Source | Citation | Section(s) |
|---|---|---|---|---|
| sra-toolkit | 3.0.7 | github.com/ncbi/sra-tools | | 3.2 |
| BBTools | 38.86 | sourceforge.net/projects/bbmap | (20) | 3.3 |
| SPAdes | 3.13.0 | github.com/ablab/spades | (21) | 3.4 |
| Bowtie2 | 2.5.1 | github.com/BenLangmead/bowtie2 | (22) | 3.5 |
| SAMtools | 1.16.1 | htslib.org | (23) | 3.5 |
| metaQUAST | 5.2.0 | github.com/ablab/quast | (24) | 3.5 |
| bioawk | 1.0 | github.com/lh3/bioawk | | 3.5 |
| MetaBAT2 | 2.15 | bitbucket.org/berkeleylab/metabat | (25) | 3.6 |
| CheckM | 1.2.2 | ecogenomics.github.io/CheckM | (26) | 3.6 |
| GTDB-Tk | 2.1.1 | github.com/Ecogenomics/GTDBTk | (27) | 3.7, 3.12 |
| geNomad | 1.7.4 | portal.nersc.gov/genomad | (28) | 3.8 |
| vRhyme | 1.1.0 | github.com/AnantharamanLab/vRhyme | (29) | 3.9, 3.12 |
| CheckV | 1.0.1 | bitbucket.org/berkeleylab/checkv | (30) | 3.9 |
| dRep | 3.5.0 | github.com/MrOlm/drep | (31) | 3.10, 3.11 |
| prodigal-gv | 2.11.0-gv | github.com/apcamargo/prodigal-gv | (28) | 3.11 |
| vConTACT2 | 0.11.3 | bitbucket.org/MAVERICLab/vcontact2 | (32) | 3.11 |
| iPHoP | 1.2.0 | bitbucket.org/srouxjgi/iphop | (33) | 3.12 |

When working with multiple software packages, as is typical in metagenomic analyses, we highly recommend the use of the package manager Conda, implemented as either Anaconda, Miniconda, or Mamba. It is not uncommon for one tool to require an exact version of a software package that conflicts with another tool. Conda allows a user to install packages to separate *environments* to avoid conflict between the dependencies of multiple tools. The installation of Conda or the individual tools used in the analyses here will not be covered, but instructions can be found by following the linked webpages for each within their cited publications.

Furthermore, workflow managers such as Snakemake (34) can play a significant role in managing complex analyses. Snakemake automates the execution of workflows, ensuring reproducibility and scalability. It allows researchers to define their analysis pipelines in a flexible, readable, and modular manner, handling dependencies and automating the execution across different computational environments. This not only saves time but also enhances the reliability of the results by reducing manual errors. The steps carried out in this chapter are treated as a ‘one-off’ analysis of metagenomes and viromes. In the real world of bioinformatics and metagenomics, it is common to refine one's methods in an iterative approach when figuring out which tools, parameters, and data are necessary to test their hypotheses. We encourage the researcher to consider workflow managers in their computational research, as they can greatly aid in the organization of computational methods and promote reproducibility of one's results.

### *2.3 - Tara Oceans Dataset*

To analyze the metagenomes of microbial and viral communities in pairs, we will leverage the publicly available *Tara* Oceans dataset (35). This dataset contains globally sampled metagenomes from 12 size fractions ranging from < 0.2 $\mu$m to > 680 $\mu$m (36). The focus of this study will be on the < 0.2 $\mu$m fraction (enrichment for viruses and other virus-like particles; hereafter viromes) and the 0.2 to 3 $\mu$m fraction (enrichment for bacteria and archaea; hereafter metagenomes). In total, there are 21 samples with both fractions available that are available to download from the NCBI Sequence Read Archive under BioProjects PRJEB1787 (metagenomes) and PRJEB4419 (viromes). This same subset of the *Tara* Oceans data was recently compared in depth to paired virome and metagenomes from other environments (37). Given the large size of the raw sequence read data for all 42 samples, this study will focus on only 10 sample pairs (20 samples total) representing the South Atlantic Gyre (35) (Table 2). Downloading the raw sequence data for these samples will be done with the NCBI SRA Toolkit (see Methods).

**Table 2. Summary of metagenomic sequence data to be analyzed.**

| Sample | Size fraction | TARA sample label | Sampling depth [m] | Environmental feature | NCBI SRA accession number |
|---|---|---|---|---|---|
| TARA_068_DCM | Metagenome | TARA_068_DCM_0.22-3 | 50 | (DCM) deep chlorophyll maximum layer | ERR599017 |
| TARA_068_DCM | Virome | TARA_068_DCM_<-0.22 | 50 | (DCM) deep chlorophyll maximum layer | ERR594415 |
| TARA_068_SRF | Metagenome | TARA_068_SRF_0.22-3 | 5 | (SRF) surface water layer | ERR599174 |
| TARA_068_SRF | Virome | TARA_068_SRF_<-0.22 | 5 | (SRF) surface water layer | ERR594391 |
| TARA_070_MES | Metagenome | TARA_070_MES_0.22-3 | 800 | (MES) mesopelagic zone | ERR599044 |
| TARA_070_MES | Virome | TARA_070_MES_<-0.22 | 800 | (MES) mesopelagic zone | ERR594407 |
| TARA_070_SRF | Metagenome | TARA_070_SRF_0.22-3 | 5 | (SRF) surface water layer | ERR599165 |
| TARA_070_SRF | Virome | TARA_070_SRF_<-0.22 | 5 | (SRF) surface water layer | ERR594353 |
| TARA_072_DCM | Metagenome | TARA_072_DCM_0.22-3 | 100 | (DCM) deep chlorophyll maximum layer | ERR599133 |

| Sample | Size fraction | TARA sample label | Sampling depth [m] | Environmental feature | NCBI SRA accession number |
|---|---|---|---|---|---|
| TARA_072_DCM | Virome | TARA_072_DCM_<-0.22 | 100 | (DCM) deep chlorophyll maximum layer | ERR594379 |
| TARA_072_MES | Metagenome | TARA_072_MES_0.22-3 | 800 | (MES) mesopelagic zone | ERR599005 |
| TARA_072_MES | Virome | TARA_072_MES_<-0.22 | 800 | (MES) mesopelagic zone | ERR594388 |
| TARA_072_SRF | Metagenome | TARA_072_SRF_0.22-3 | 5 | (SRF) surface water layer | ERR598984 |
| TARA_072_SRF | Virome | TARA_072_SRF_<-0.22 | 5 | (SRF) surface water layer | ERR594364 |
| TARA_076_DCM | Metagenome | TARA_076_DCM_0.22-3 | 150 | (DCM) deep chlorophyll maximum layer | ERR599148 |
| TARA_076_DCM | Virome | TARA_076_DCM_<-0.22 | 150 | (DCM) deep chlorophyll maximum layer | ERR594355 |
| TARA_076_SRF | Metagenome | TARA_076_SRF_0.22-3 | 5 | (SRF) surface water layer | ERR599126 |
| TARA_076_SRF | Virome | TARA_076_SRF_<-0.22 | 5 | (SRF) surface water layer | ERR594354 |
| TARA_078_SRF | Metagenome | TARA_078_SRF_0.22-3 | 5 | (SRF) surface water layer | ERR599006 |
| TARA_078_SRF | Virome | TARA_078_SRF_<-0.22 | 5 | (SRF) surface water layer | ERR594411 |

# 3 - Methods

## *3.1 - Study Design*

There are various factors that one can consider when designing a metagenomics-based microbiome study. The nuances in the design of a project will depend on one's study system and the individual research questions being asked. Thus, study design and sampling methods can have a significant impact on one's interpretation of the ecology of microbiomes (37, 38). For any investigation into microbial and viral metagenomics, we suggest three key factors for the researcher to consider when designing their study. These three factors will also dictate the hardware requirements to analyze the resulting sequence data (see Materials), so each factor should be carefully considered.

The first factor one should consider is the scale of the microbiome study, i.e. the variation of biological and technical replicates. We recommend prioritizing biological replicates over technical replicates. This is because biological replicates can improve the power and accuracy of sequence-based analyses better than technical replicates (39). Second, it should be considered whether to enrich for any biological groups. For example, any microbiome study with an interest in viruses should generate virus-sized fraction metagenome (viromes) from their samples (37, 40). We argue that any holistic investigation into a microbial community should consider viruses in addition to the microbes that they infect. Third, the target sequencing depth of each biological sample is crucial. The more DNA sequence data generated per sample, the greater the resolution will be for fine-level variation of genes and genomes (41). Greater sequencing depth will also allow for the detection of rare members of a community (42, 43). However, increasing the desired sequencing depth will also increase the cost of the study. This presents a trade-off between the financial costs of increasing sequencing depth and information able to be generated with increased depth (44).

There are virtually endless microbiomes not yet investigated or sequenced on Earth with an equally long list of questions that one could ask about them. Yet, there is already a vast amount of publicly available microbiome data. These data are stored in digital repositories such as NCBI GenBank, JGI IMG, and EMBL-EBI ENA, among others. Before embarking on a metagenomic sequencing project, small or large, we encourage the researcher to ask themselves if their hypotheses can be tested using already publicly available sequencing data. In doing so, it is the responsibility of the researcher to obtain the consent of those who originally generated the data. One resource that was explicitly generated for the widespread use of the scientific public is the *Tara* Oceans dataset (35), which will be used for the analyses here (see Materials). This dataset involves a study design that expands time (sample collection date), spatial (geography and water column depth), and physiochemical (water column depth and environmental feature) scales. The dataset also includes the enrichment of multiple biological entities such as microbes and viruses. Thus, it is well suited for multiple in-depth investigations into whole microbial and viral communities of microbiomes in the environment.

### *3.2 - Accessing Publicly Available, Raw Metagenomic Sequence Reads*

The subset of *Tara* data used in this analysis (see Materials) will be downloaded using the `prefetch` command from the NCBI SRA Toolkit, followed by `fasterq-dump` to produce raw sequence read files in `.fastq` format. To download multiple `.fastq` files in bulk, a list of accession numbers can be provided. A simple text file can be created and edited using a command-line text editor like `vim` or `nano` which usually come pre-installed in many operating systems. A column of accession numbers such as that in Table 2 can be copied from a local spreadsheet and pasted into the text editor. Using the `cat` command, we show the contents of this file below.

```
cat acc_list.txt

ERR599017
ERR594415
ERR599174
ERR594391
ERR599044
ERR594407
ERR599165
ERR594353
ERR599133
ERR594379
```

```
ERR599005
ERR594388
ERR598984
ERR594364
ERR599148
ERR594355
ERR599126
ERR594354
ERR599006
ERR594411
```

The file `acc_list.txt` can now be provided to `prefetch`. The flag `–max-size 50g` is added because some of the files will be larger than the default maximum file size of 20 GB.

```
prefetch --option-file acc_list.txt --output-directory prefetch_out --max-siz
e 50g
```

The folder `prefetch_out` now contains one folder for each accession in the list. Within each folder are `.sra` files that will be read by `fasterq-dump`.

```
fasterq-dump prefetch_out/ERR* --outdir raw_reads

spots read       : 47,220,407
reads read       : 94,440,814
reads written    : 94,440,814
# (Remaining output truncated)
```

The '*' following `prefetch_out/ERR` is known as a *wildcard* character. Using it here returned all files or folders under `prefetch_out` that have the prefix `ERR` so that each individual folder did not have to be typed in the `fasterq-dump` command. There should now be a new folder named `raw_reads` that contains two `.fastq` files per sample (`SAMPLENAME_1.fastq`, `SAMPLENAME_2.fastq`). That is because these are *paired-end* sequence reads. Each sample has a forward (`_1`) and a reverse (`_2`) set of reads (a *mate*) that are used together as a pair. After the paired-end `.fastq` files have been downloaded into `raw_reads`, the `prefetch_out` folder can be removed to save on storage using the command `rm -r prefetch_out`.

To further save on disk space, each of the generated `.fastq` files will be compressed using the command `pigz`. The wildcard character '*' will be used again, this time to specify all files under `raw_reads` that end with the file extension `.fastq`. This will convert all `.fastq` files in the `raw_reads` folder to compressed files with the extension `.fastq.gz`. Many bioinformatics tools that handle sequence reads can conveniently handle `.fastq.gz` files instead of the large `.fastq` files.

```
pigz raw_reads/*.fastq
```

### *3.3 - Sequence Read Quality Control (QC)*

The most important step in any metagenomics project involves processing and QC of the raw sequenced reads obtained directly from a sequencing center or data repository. As the old saying goes, *garbage in, garbage out*. Even with highly accurate next generation sequencing technologies, raw sequence read libraries contain poor quality and/or contaminating reads that need to be removed. There have been multiple tools developed to address this issue. Some of the more popular and robust tools include (but are not limited to) FastQC for initial read quality assessment (45), Trimmomatic for the removal of low-quality bases in Illumina paired end reads (46), and BBDuk for decontamination (20). One should consider using a tool like FastQC to assess the quality of reads before QC and again on reads after QC. Comparing the two reports can reveal how one's choice of QC methods improved (or perhaps did not improve) sequence read quality.

The rqcfilter2 program from BBTools (20) is a package that will be used here to perform read adapter removal, read trimming, contaminant filtering, and read quality filtering all in one execution. It requires an associated database that is 106 GB in size. A *for-loop* can be executed on the command-line using user-defined variables `DATADIR`, `THREADS`, `MEM`, and `OUT` to set the location of the RQCFilter data, maximum-allowable CPU usage, maximum-allowable memory usage, and the output folder for the filtered reads. Within the for-loop, the IDs from each mate in the pair of `.fastq.gz` files can be inferred using bash string manipulation and wildcards so that they can be dynamically provided to `rqcfilter2.sh`, instead of manually providing them one-by-one. Using `nano` or `vim` again, the code shown below can be written to a file named `qc_reads.sh`:

```
#!/bin/sh
# qc_reads.sh

DATADIR="RQCFilterData"
THREADS="16"
MEM="64"
OUT="filtered_reads"

for FASTQ in raw_reads/*_1.fastq.gz; do
        BASE=$(basename "$FASTQ")
        # Extract the sample name
        SAMPLE=$(echo "$BASE" | cut -d'_' -f1)
        F="$FASTQ"
        # Extract the reverse mate filename
        R=$(echo "$F" | sed 's/_1.fastq.gz/_2.fastq.gz/')

        echo "Running rqcfilter2.sh (BBTools) on $F $R"
        rqcfilter2.sh \
                in="$F" in2="$R" \
                rqcfilterdata="$DATADIR" \
                trimfragadapter=t \
                qtrim=r trimq=0 \
                maxns=3 maq=3 \
                minlen=51 mlf=0.33 \
                phix=t removehuman=t \
```

```
                removedog=t removecat=t \
                removemouse=t removemicrobes=f \
                clumpify=t tmpdir= \
                barcodefilter=f trimpolyg=5 \
                usejni=f chastityfilter=f \
                tmpdir=rqcfilter2_tmp \
                path="$OUT" \
                out="${SAMPLE}_1.fastq.gz" \
                out2="${SAMPLE}_2.fastq.gz" \
                threads="$THREADS" -Xmx"$MEM"g

        rm -r rqcfilter2_tmp
        echo "Done $SAMPLE."
done
```

The flags `phix=t removehuman=t removedog=t removecat=t removemouse=t` removes contaminants by mapping the reads to phage Phi X (commonly added to sequencing runs), human, dog, cat, and mouse reference sequences in the `RQCFilterData` database, while the flag `removemicrobes=f` ensures that reads mapping to microbial reference sequences are retained. The remaining flags such as `trimfragadapter=t qtrim=r trimq=0 maxns=3 maq=3 minlen=51 mlf=0.33` are quality trimming parameters that the user can adjust. These values were chosen to mimic the parameters used in the JGI Metagenome Workflow (17). The remaining flags `clumpify=t tmpdir= barcodefilter=f trimpolyg=5 usejni=f chastityfilter=f` are additional flags that the user may or may not wish to enable depending on the sequencing technology used and how the reads were obtained. There are many parameters that can be changed when running `rqcfilter2.sh`, we encourage the user to review and consider them with `rqcfilter2.sh -help`.

This may take some time to run, so to execute this script in the background the `nohup` command can be used with the addition of a ‘`&`‘. The text that would have been printed to the user’s screen can be saved in a file using a ‘`>`‘ operator a specifying an output file for the screen text:

```
nohup sh qc_reads.sh > qc_reads.nohup.out &
```

After the for-loop is completed, the contents of `qc_reads.nohup.out` can be viewed with the `less` command to see what printed to the ‘screen’ when running the script, if desired.

The folder `filtered_reads` now contains the filtered mates produced by `rqcfilter2.sh`. Although not required for genome assembly, error correcting filtered reads before they are assembled into a metagenome can greatly improve assembly quality (47, 48). However, it should be noted that error correction may not always improve assembly quality. Error correction typically is not beneficial when sequencing depth is low (47, 48) and results can vary depending on the error-correction software and/or assembler used (47, 48). Judging by the file sizes alone, the sequence reads in the *Tara* dataset here were sequenced deeply. Additionally, SPAdes (21) the assembly workflow to be applied below (Section 3.4) performs optimally with read error correction. There are several tools available for read error correction and their applications vary by sequencing technology (47). Below, we demonstrate how to error-correct filtered reads using yet another tool from the BBTools suite, `bbcms.sh`. Just as what was done to run `rqcfilter2.sh`, another bash script, `correct_reads.sh`, can execute this command for each pair of filtered reads.

```
#!/bin/sh
# correct_reads.sh

THREADS="16"
MEM="64"
OUT="corrected_reads"

for FASTQ in filtered_reads/*_1.fastq.gz; do
        BASE=$(basename "$FASTQ")
        # Extract the sample name
        SAMPLE=$(echo "$BASE" | cut -d'_' -f1)
        F="$FASTQ"
        # Extract the reverse mate filename
        R=$(echo "$F" | sed 's/_1.fastq.gz/_2.fastq.gz/')
        # Set the output path for forward corrected reads
        OUTF="${OUT}/${SAMPLE}_1.fastq.gz"
        # Set the output path for reverse corrected reads
        OUTR="${OUT}/${SAMPLE}_2.fastq.gz"

        echo "Running bbcms.sh (BBTools) on $F $R"
        bbcms.sh \
            -Xmx"$MEM"g \
            t="$THREADS" \
            mincount=2 \
            highcountfraction=0.6 \
            in="$F" \
            in2="$R" \
            out="$OUTF" \
            out2="$OUTR"

        echo "Done $SAMPLE."
done
```

The options `mincount=2 highcountfraction=0.6` are specified to discard reads largely composed of very rare k-mers, which can lead to issue in assembly downstream. These options were chosen to mimic the implementation of `bbcms.sh` in the JGI Metagenome Workflow (17) to adhere to standardization. However, the user should consider all options available with bbcms.sh and set the parameters that are most suited for their data.

The folder `corrected_reads` now exists and contains the error corrected, filtered reads. However, during the QC and error correction process, it is not uncommon for one mate in the pair to have been discarded while the other did not. This results in an unequal number between the forward and reverse reads for each sample. Genome and metagenome assemblers typically do not allow this, so the unpaired (singleton) reads will have to be separated into their own file. This can be done with another tool in the BBTools suite, `repair.sh`, which will be executed in another for-loop in the script below named `repair_reads.sh`.

```
#!/bin/sh
# repair_reads.sh

DATADIR="RQCFilterData"
MEM="64"
OUT="corrected_reads"

for FASTQ in corrected_reads/*_1.fastq.gz; do
    BASE=$(basename "$FASTQ")
    # Extract the sample name
    SAMPLE=$(echo "$BASE" | cut -d'_' -f1)
    F="$FASTQ"
    # Extract the reverse mate filename
    R=$(echo "$F" | sed 's/_1.fastq.gz/_2.fastq.gz/')
    # Set the output path for forward repaired reads
    FILTOUTF="${OUT}/${SAMPLE}_F.fastq.gz"
    # Set the output path for reverse repaired reads
    FILTOUTR="${OUT}/${SAMPLE}_R.fastq.gz"
    # Set the unpaired reads filename
    FILTOUTS="${OUT}/${SAMPLE}_U.fastq.gz"

    # Separate filtered reads into F, R, and U reads
    echo "Running repair.sh (BBTools) on $FILTOUT"
    repair.sh \
        -Xmx"$MEM"g \
        in="$F" in2="$R" \
        out="$FILTOUTF" \
        out2="$FILTOUTR" \
        outs="$FILTOUTS" \
        repair

    echo "Done $SAMPLE."
done
```

Now, all of the raw reads have been properly QC'd and error corrected. The non-'repaired' and the non-error-corrected reads can be removed with the following command:

```
rm -r corrected_reads/*_1.fastq.gz corrected_reads/*_2.fastq.gz filtered_read
s
```

Since the raw reads were obtained from an online archive, they can be removed as well to save on space with `rm -r raw_reads`. However, the original raw reads for every sequencing project should always be saved or archived. If reads were filtered from a collection of raw reads that have not yet been published or archived, the raw reads should *not* be removed.

### *3.4 - Metagenome Assembly*

The assembly of reads is the nucleus of any metagenomics project. Given the high complexity of metagenomic sequence data, even after QC and error correction, it is imperative to pick an assembler that is most suited for the data. Not all assemblers are able to assemble metagenomes and several assemblers are designed for a particular sequencing technology (49–51). Paired-end, next-generation (i.e. *Illumina*) sequencing is the most common technology used for metagenomes, so it may be unsurprising that most modern assemblers are designed for these reads. While several genome and metagenome assemblers have been developed over the years and continue to be developed to this day, two assemblers have proven to be particularly reliable. SPAdes (21) and MEGAHIT (51) are paired end read assemblers designed for single genomes and metagenomes. They are both extensively documented and maintained which makes them quite user friendly. Neither tool consistently performs better than the other, their performance varies from dataset to dataset (43, 52). However, the metagenome assembly mode of SPAdes (metaSPAdes) requires substantially more RAM than MEGAHIT (52). Since the performance of each tool varies on a case-by-case basis, we recommend that the researcher assembles their metagenomics reads with multiple assemblers and then compare the results (see 3.5 Metagenome QC). For the purposes of this guide, we will only demonstrate how to assemble the error corrected metagenomic reads with metaSPAdes.

metaSPAdes can be run on every set of error corrected reads using another shell script with a for-loop, `assemble.sh`.

```
#!/bin/sh
# assemble.sh

THREADS="16"
MEM="100"
OUT="assemblies"

mkdir -p $OUT

for FASTQ in corrected_reads/*_F.fastq.gz; do
    BASE=$(basename "$FASTQ")
    # Extract the sample name
    SAMPLE=$(echo "$BASE" | cut -d'_' -f1)
    F="$FASTQ"
    # Extract the reverse mate filename
    R=$(echo "$F" | sed 's/_F.fastq.gz/_R.fastq.gz/')
    # Extract the unpaired reads filename
    U=$(echo "$F" | sed 's/_F.fastq.gz/_U.fastq.gz/')

    echo "Running metaSPAdes to assemble reads $F $R $U"
    spades.py \
        -1 "$F" -2 "$R" -s "$U" \
        -o "$OUT"/"$SAMPLE" \
        --meta \
```

```
        --only-assembler \
        --tmp-dir spades_tmp \
        -t "$THREADS" -m "$MEM"

    rm -r spades_tmp
    echo "Done $SAMPLE."
done
```

Here the `-meta` flag enables metagenome assemble mode. The flag `-only-assembler` disables error correction since this was already done with `bbcms.sh`, above. Assuming this script was run in the background and had the screen output directed to a text file with `nohup sh assemble.sh > assemble.nohup.out &`, one can quickly check how many SPAdes runs were successful by counting the occurrences of the phrase '`SPAdes pipeline finished`' in the screen output. This is because SPAdes will print this message to the screen after a successful assembly. Of course, counting the occurrences of this phrase will not apply to other assemblers, but the user can `grep -c` another phrase that their assembler of choice may print only when assembly has finished.

```
grep -c "SPAdes pipeline finished" assemble.nohup.out

20
```

If any of the runs failed to complete because the 100 GB memory limit was exceeded, and no more memory is available to use, attempting assembly again with MEGAHIT will be a good option.

The folder `assemblies` should now contain one sub-folder for each *Tara* sample assembled. Inside of each sub-folder is a file `contigs.fasta` that contains reads that were assembled into *contigs*. A contig is a contiguous stretch of an assembled DNA sequence. The paired end sequencing technology used allowed for the contigs to be joined by SPAdes into *scaffolds* in the file `scaffolds.fasta`. Scaffolds are essentially multiple contigs that were joined by a sequence of Ns. The Ns represent unknown nucleotides. Here, the number of Ns in a sequence correspond to the number of base pairs between the stretches of DNA represented by the contigs. In essence, having a set of contigs that were joined into a scaffold means that the orientation of each contig relative to each other is known, the distance (base pairs) between the contigs is known, but the exact nucleotides that the Ns represent are unknown. The advantage of scaffolds is that they allow for greater contiguity of an assembly, which can help retain genomic context of encoded genes. If one prefers to analyze sequences where every base is known, then one can choose to continue their analysis with the contigs instead of scaffolds. Here, the assembled scaffolds will be used for further analysis.

A new folder can be made to store just the assembled scaffolds with `mkdir scaffolds`. Next, the scaffold files can be copied to the new folder, each file renamed with their original sample identifiers, with the following code:

```
find assemblies -mindepth 2 -maxdepth 2 -type f -name "scaffolds.fasta" -exec
sh -c 'cp {} scaffolds/$(basename $(dirname {}))_scaffolds.fasta' \;
```

The contents of the scaffolds folder can be inspected with ls scaffolds. The folder should not contain the following:

```
ERR594353_scaffolds.fasta
ERR594354_scaffolds.fasta
ERR594355_scaffolds.fasta
ERR594364_scaffolds.fasta
ERR594379_scaffolds.fasta
ERR594388_scaffolds.fasta
ERR594391_scaffolds.fasta
ERR594407_scaffolds.fasta
ERR594411_scaffolds.fasta
ERR594415_scaffolds.fasta
ERR598984_scaffolds.fasta
ERR599005_scaffolds.fasta
ERR599006_scaffolds.fasta
ERR599017_scaffolds.fasta
ERR599044_scaffolds.fasta
ERR599126_scaffolds.fasta
ERR599133_scaffolds.fasta
ERR599148_scaffolds.fasta
ERR599165_scaffolds.fasta
ERR599174_scaffolds.fasta
```

Having each scaffolds file be named by their sample origin will be useful when comparing assemblies to each other. The remaining SPAdes output files will not be used hereafter. They will be archived using the `tar` command with compression using `pigz`, and the original folder will be removed to save on disk space:

```
tar --use-compress-program="pigz --best --recursive" -cf assemblies.tar.gz as
semblies && rm -r assemblies
```

The `&&` simply means that when the first part of the code finishes running without error, the second part `rm -r assemblies` will be executed.

### *3.5 - Metagenome QC*

Just as QC was performed on sequence reads before assembly, the assembled metagenome scaffolds will also need to undergo QC. SPAdes successfully ran for each of the 20 samples, but it is not clear at this point what fraction of sequence reads ended up in the final assembly, how many scaffolds were assembled in each sample, how long they are, or what fraction of the scaffolds are composed of Ns. These are each important indicators of the overall quality of a (meta)genome assembly.

An ideal metagenome will have been assembled from all input reads, but in reality, not all input reads will end up in the final assembly. The *read recruitment* of scaffolds is a good indicator of how much genomic information in samples went unobserved. Read recruitment is the percentage of reads that went into an assembler that end up mapping back to the output scaffolds. Mapping the filtered, error-corrected reads to the scaffolds will allow for the read recruitment to be calculated. A higher read recruitment of a metagenome assembly is indicative of a higher overall metagenome quality. In other words, a metagenome with a read recruitment of 90% is a better representative of the actual sequenced microbial community than a metagenome with a read recruitment of 25%. Unless deliberate measures were taken to filter the scaffolds resulting from assembly, the assemblies should recruit at least 50% of the reads.

The tool Bowtie2 (22) will be used to map the filtered, error-corrected reads to the scaffolds. In addition to calculating read recruitment, mapping reads this way will be useful for further analyses below. Bowtie2 is a great read-mapper for short-read data. For metagenomes assembled from long reads, minimap2 (53) is recommended. First, the scaffolds need to be *indexed* by Bowtie2. This will be done using a short shell script `bowtie_index.sh` with a for-loop, sending the output indices to a new folder `bowtie2_index`:

```
#!/bin/sh
# bowtie_index.sh

THREADS="16"
for FASTA in scaffolds/*_scaffolds.fasta; do
    bowtie2-build "$FASTA" \
    "bowtie2_index/$(basename "${FASTA%_scaffolds.fasta}")" \
    --threads "$THREADS"
done
```

Now each set of reads that went into metagenome assembly can be mapped to their respective indices with the `bowtie2` command using another shell script `bowtie_map.sh`. The screen output from the Bowtie2 runs will be used below. To save the screen output and run Bowtie2 in the background, the following script will be executed with `nohup sh bowtie_map.sh > bowtie_map.nohup.out &`:

```
#!/bin/sh
# bowtie_map.sh

INDEX_DIR="bowtie2_index"
READS_DIR="corrected_reads"
OUTPUT_DIR="map_files"
THREADS="16"

mkdir -p "$OUTPUT_DIR"

for FWD_READ in ${READS_DIR}/*_F.fastq.gz; do
    # Extract the base name without the forward suffix
    BASE_NAME=$(basename "$FWD_READ" _F.fastq.gz)

    # Specify the matching reverse read
    REV_READ="${READS_DIR}/${BASE_NAME}_R.fastq.gz"
    # Specify the singleton read
    UNPAIRED_READ="${READS_DIR}/${BASE_NAME}_U.fastq.gz"
```

```
    # Run bowtie2 with paired-end and unpaired (singleton) reads
    bowtie2 -x "${INDEX_DIR}/${BASE_NAME}" \
        -1 "$FWD_READ" -2 "$REV_READ" \
        -U "$UNPAIRED_READ" \
        -S "${OUTPUT_DIR}/${BASE_NAME}.sam" \
        --threads "$THREADS"

    # Convert SAM to BAM, sort, and index
    samtools view \
        -bS "${OUTPUT_DIR}/${BASE_NAME}.sam" \
        --threads "$THREADS" \
        | samtools sort \
            -o "${OUTPUT_DIR}/${BASE_NAME}.bam" \
            --threads "$THREADS"
    rm "${OUTPUT_DIR}/${BASE_NAME}.sam"
    samtools index "${OUTPUT_DIR}/${BASE_NAME}.bam" -@ "$THREADS"
done
```

There are four things that the above script does:

1. The command `bowtie2` is executed for each sample among the reads and indices. The flag `-x` specifies the target index, while `-1`, `-2`, and `-U` specify the forward, reverse, and unpaired reads for that same sample. The flag `-S` writes the output alignment file in SAM format. The output SAM file is a large, human-readable file with mapping statistics for every read.
2. Since SAM files are seldom used on their own, the SAM files are converted in BAM format to save on space using the `samtools view -bS` command from SAMtools (23). BAM is a binary version of SAM. It contains the same information but in binary rather than human-readable text.
3. The BAM-formatted output is provided as an input (using the *pipe* operator `|`) to `samtools sort` which will alpha-numerically sort the information before writing the final BAM file using the sample name and the `.bam` extension. Sorting BAM files this way is a common prerequisite for many tools that handle them. The SAM-formatted file is removed since it is now no longer needed.
4. The sorted BAM file is indexed. Indexing BAM files is another common prerequisite for tools that handle them.

Now that the reads that went into metagenome assembly have been mapped to the filtered scaffolds, read recruitment can be assessed. Read recruitment can be calculated at any point after read mapping using tools such as CoverM (github.com/wwood/CoverM). However, Bowtie2 automatically reports the read recruitment to each index after mapping in its screen output. To obtain these values from the `bowtie_map.nohup.out` file, the following can be executed:

```
grep "overall alignment rate" bowtie_map.nohup.out

89.33% overall alignment rate
91.36% overall alignment rate
88.37% overall alignment rate
```

```
88.65% overall alignment rate
90.32% overall alignment rate
86.22% overall alignment rate
89.64% overall alignment rate
92.86% overall alignment rate
87.90% overall alignment rate
88.61% overall alignment rate
83.33% overall alignment rate
78.23% overall alignment rate
73.42% overall alignment rate
73.52% overall alignment rate
89.11% overall alignment rate
50.99% overall alignment rate
85.38% overall alignment rate
73.98% overall alignment rate
73.57% overall alignment rate
77.73% overall alignment rate
```

Every index built from the SPAdes scaffolds recruited at least 50% of the reads, and several much more than that. This is a good sign that the previous sequence read QC and subsequent assembly methods worked as intended. However, read recruitment is not the only way to measure assembly quality and perform metagenome QC. Another important factor to consider is the contiguity of an assembly. An ideal metagenome assembly would have one scaffold per genetic element (chromosome, plasmid, virus, etc.) that existed in the actual sample. In reality, these elements often appear highly fragmented in metagenomes due the deliberate fragmentation of nucleic acids during DNA sequencing (an approach known as 'shotgun' sequencing) and from the DNA extraction process itself (54). Greater fragmentation can lead to issues with metagenome binning and sequence analysis later (54). The magnitude of this problem can typically be reduced with greater sequencing depth (54) or with high-fidelity long-read sequencing (54), but these options are not always financially or technically viable.

To measure the contiguity of the metagenomes, the tool metaQUAST (24) will be used. metaQUAST will calculate several summary statistics to assess metagenome quality and provide tabular as well as interactive, graphical outputs. Here, metaQUAST is used because of its speed when calculating basic statistics. Below is an example of a basic metaQUAST command:

```
metaquast scaffolds/* --min-contig 1000 --max-ref-number 0 --threads 16 --out
put-dir metaquast
```

Using `scaffolds/*` after the `metaquast` command will run the metaQUAST workflow on all scaffolds in the `scaffolds` folder. The flag `–min-contig 1000` sets the minimum contig (in this case, scaffold) length for analysis to 1000 bp. The flag `–max-ref-number 0` will skip downloading reference genomes and comparing to the input assemblies, which will reduce runtime. The maximum number of threads to use is specified with `–threads 16` and the output folder '`metaquast`' is specified with `–output-dir metaquast`.

A table of summary statistics for each input metagenome will be provided in the output `report.tsv`. An interactive report `report.html` will also be generated, which can be opened in a web browser. This interactive report is very powerful and will provide many different visualizations of metagenome assembly statistics. There are several tables and graphs generated by metaQUAST that are useful to consider when performing metagenome QC. One of the most useful reports is a table with length and contiguity statistics for every input metagenome (only six assemblies shown, see Figure 1).

**QUAST**

Quality Assessment Tool for Genome Assemblies by CAB

28 February 2024, Wednesday, 09:31:36

View in Icarus contig browser

All statistics are based on contigs of size >= 1000 bp, unless otherwise noted (e.g., "# contigs (>= 0 bp)" and "Total length (>= 0 bp)" include all contigs).

Worst Median Best ☑ Show heatmap

| Statistics without reference | ERR599017_scaffolds | ERR594415_scaffolds | ERR599174_scaffolds | ERR594391_scaffolds | ERR599044_scaffolds | ERR594407_scaffolds |
|---|---|---|---|---|---|---|
| # contigs | 57 644 | 52 382 | 63 593 | 62 195 | 89 318 | 24 038 |
| # contigs (>= 0 bp) | 591 973 | 241 357 | 636 346 | 278 064 | 2 444 899 | 142 226 |
| # contigs (>= 1000 bp) | 57 644 | 52 382 | 63 593 | 62 195 | 89 318 | 24 038 |
| # contigs (>= 5000 bp) | 3245 | 6229 | 3557 | 8886 | 8412 | 4057 |
| # contigs (>= 10000 bp) | 592 | 2293 | 699 | 3372 | 3297 | 1995 |
| # contigs (>= 25000 bp) | 32 | 508 | 49 | 850 | 920 | 642 |
| # contigs (>= 50000 bp) | 1 | 102 | 3 | 234 | 311 | 265 |
| Largest contig | 55 361 | 179 113 | 83 971 | 1 693 895 | 579 666 | 1 456 704 |
| Total length | 124 160 796 | 163 041 179 | 136 780 541 | 230 111 134 | 268 616 874 | 133 149 004 |
| Total length (>= 0 bp) | 342 737 291 | 249 413 992 | 373 173 543 | 326 789 782 | 859 304 933 | 184 723 123 |
| Total length (>= 1000 bp) | 124 160 796 | 163 041 179 | 136 780 541 | 230 111 134 | 268 616 874 | 133 149 004 |
| Total length (>= 5000 bp) | 26 062 594 | 74 408 487 | 29 619 535 | 125 279 743 | 123 941 686 | 95 604 064 |
| Total length (>= 10000 bp) | 8 535 172 | 47 437 714 | 10 444 199 | 87 457 523 | 89 087 341 | 81 230 736 |
| Total length (>= 25000 bp) | 984 448 | 20 840 835 | 1 638 237 | 49 922 511 | 53 457 785 | 60 619 367 |
| Total length (>= 50000 bp) | 55 361 | 7 108 834 | 193 303 | 29 281 462 | 32 369 796 | 47 737 913 |
| N50 | 2321 | 4257 | 2296 | 5946 | 4206 | 19 200 |
| N90 | 1152 | 1295 | 1149 | 1397 | 1224 | 1603 |
| auN | 3831.9 | 11 374 | 4049 | 52 178 | 25 632 | 167 085 |
| L50 | 14 439 | 7774 | 15 685 | 7013 | 10 676 | 917 |
| L90 | 46 052 | 38 007 | 50 814 | 42 612 | 64 924 | 13 362 |
| GC (%) | 43.97 | 37.12 | 44.98 | 44.04 | 48.67 | 55.07 |
| **Mismatches** | | | | | | |
| # N's per 100 kbp | 0 | 119.08 | 0 | 143.08 | 62.22 | 111.23 |
| # N's | 0 | 194 151 | 0 | 329 239 | 167 133 | 148 102 |

**Figure 1: A Subset of metaQUAST Statistics for Assembled Metagenomes.** Table captured from the metaQUAST html output.

The number of contigs (again, scaffolds in this case) is the first reported statistic. The number of scaffolds in an assembly is not a sufficient statistic by itself to assess metagenome quality. For example, sample ERR599044 has more scaffolds than ERR594407, but the former is composed of shorter scaffolds than the latter. How is this known? For one, the largest scaffold in ERR599044 is reported to be 580 kb while the largest in ERR594407 is 1.46 Mb. Though, the most informative statistics here are N50, L50, N90, and L90. These four statistics are direct measurements of contiguity.

N50 is defined as "the length of the shortest contig for which longer and equal length contigs cover at least 50% of the assembly" (55). In other words, if one were to order all scaffolds in an assembly by decreasing length, the N50 of that assembly would be the size of the scaffold in the very middle of that list. Here, ERR599017 has a lower N50 than ERR599044, suggesting that the former is less contiguous than the latter. N90 is like N50, except the value gives the length of the contig that is longer than 90% of an assembly.

L50 and L90 are also like N50 and N90, but instead of the values representing sequence lengths, they represent the rank order of a scaffold by length. In other words, the L50 of 7,774 for ERR594415 means that 7,774 scaffolds make up 50% of the length of the entire metagenome. Lower L50/L90 values are indicative of greater contiguity. Additionally, metaQUAST has counted the number of Ns in the scaffolds. In some cases, more Ns are met with greater contiguity because of scaffolding. For example, ERR594407 has 79% more Ns per 100 kb than ERR599044, but is has a considerably higher N50 and lower L50. While a smaller fraction of Ns in a metagenome assembly is desired, the fact that Ns are present may be indicative that genomic structure was preserved during assembly, maintaining contiguity.

So, what to the reported values here mean for quality control of the assembled *Tara* metagenomes? Generally, more contiguous metagenomes are of a higher quality than discontinuous metagenomes. However, it is not easy to provide a specific N50/N90 or L50/L90 value that should be used as a cutoff to determine whether a metagenome should be analyzed or not. This is especially the case with the dataset being used here because there is a biological reason to only compare viromes to viromes and metagenomes to metagenomes. Half of the assemblies are viromes (shown here are ERR599017, ERR599174, and ERR599044), representing the virus-sized fraction of a sample. Viruses have genome sizes orders of magnitude smaller than that of their cellular hosts, so it makes biological sense for a virome to appear more discontinuous than a cellular metagenome. In this case, it may not necessarily be true that the viral genomes are more fractured than the total metagenomes, but instead they are just smaller genomes. Additionally, without knowing what the true community composition is in each sample, it is improbable to come up with a desired N- or L- statistic for any given sample since there will be real biological variation in genome composition within samples. Having prior knowledge of sampling design such as that in Table 2 is useful here. If it is assumed that biological and or methodological replicates should have similar microbial communities, by extension they should have similar N- or L-statistics and scaffold lengths. Assemblies within these replicates that deviate from the rest could be candidates for removal or further curation.

It is generally recommended to remove scaffolds < 1,000 bp from metagenome assemblies before genome binning and further analysis. Retaining scaffolds < 1,000 bp in metagenomes can lead to increased computation times for downstream analyses with little informational reward (2, 43). They can also make binning difficult (56, 57). Therefore, removing scaffolds < 1,000 bp is often beneficial and a good step to perform in metagenome QC. Doing so with the assembled *Tara* metagenomes here will result in the removal of a significant fraction of all base pairs in each assembly. This is sometimes the unkind reality of metagenomics. One may assume that all of the DNA in an environmental sample may be chromosomal or otherwise circular and organized. However, environmental DNA, protected or not, may come from a wide variety of unknown sources and can end up making up a large fraction of a final DNA sample (58, 59).

Highly fragmented assemblies like these can sometimes be mitigated by increasing sequencing depth (54). Unless there is prior information available about the community composition of a sample, it is difficult to estimate the ideal sequencing depth to result in a more contiguous metagenome assembly (60). Additionally, metagenome co-assembly can sometimes improve contiguity. Co-assembly is the process of combining separate metagenomic read samples (ideally technical replicates) and assembling them together. On the one hand, the added read depth of similar microbial communities can make assembling rare and variable community members more successful (61). On the other hand, if there is a lot of inter-sample variation, co-assembly may not end up improving assembly quality due to the added complexity (31). Here, not all depths and locations are replicated across the 20 *Tara* samples, and viromes and metagenomes should not be coassembled. Therefore, the single-assembly approach will be upheld to avoid biasing assembly quality towards any sample location or depth.

In any case, unless there is a clear biological reason to retain very short sequences, one can use a single line of code to achieve this using `bioawk`:

```
mkdir -p filtered_scaffolds

for FASTA in scaffolds/*_scaffolds.fasta; do \
  bioawk -c fastx '{ if(length($seq) >= 1000) \
  print ">"$name"\n"$seq }' $FASTA \
  > filtered_scaffolds/$(basename $FASTA); done
```

The filtered scaffolds will be in a new folder `filtered_scaffolds`. Comparing the sizes of the two folders with `du -sh scaffolds filtered_scaffolds` shows volumes of 13 GB and 4 GB, a near 70% reduction in file size. That is a lot of low-quality data removed! This should not impact the statistics calculated by metaQUAST since it was already set to only analyze sequences >= 1,000 bp with `-min-contig 1000`. On the other hand, read recruitment will surely be affected by removing 70% of the bases across all assemblies. Just as read recruitment was assessed for the unfiltered scaffolds, it can be done again using the filtered scaffolds to see how much of the original sequence data ends up in the filtered assemblies. The binning process below requires BAM files for reads mapped to the filtered scaffolds. It will be assumed that the researcher has generated these and put them in a new folder named `map_files_filt`. To be concise, this will not be shown, but once can expect a drop in read recruitment considering the number of scaffolds removed.

### *3.6 - Binning Microbial Genomes*

The fundamental difference between metagenome assembly and single-genome assembly is that metagenomes contain DNA fragments of multiple community members. Metagenomics was first used to extract meaningful information (gene clusters) from uncultivated microbes in a community without genome-level resolution (1). It later became possible to reconstruct whole genomes from metagenomes *de novo* (2, 3), which allows for a better understanding of microbial populations in their original environments. Generating *metagenome-assembled genomes* (MAGs) is therefore a powerful step in nearly any metagenomic study. Generating MAGs, a process not unlike to putting together pieces of a puzzle, is called *binning*. There have been a variety of tools developed to bin microbial scaffolds into MAGs (25, 62–65). The most common ones used today are MetaBAT2 (25), MaxBin2 (62), and CONCOCT (65). These three binning tools are incorporated into the wrapper called metaWRAP (15), which allows users to bin their scaffolds with all three tools and then refine them into consensus bins. While metaWRAP's multi-tool and refinement approach can be useful, metaBAT2 alone often performs very well and it is much simpler to run. This is especially the case when mapped read `.bam` files are provided to assist in the binning process. Here, metaBAT2 will be used to bin the filtered mixed-community metagenome scaffolds into MAGs. Scaffolds originating from the viromes will be binned with a different approach below (see Section 3.9).

To run metaBAT2 on every mixed-community metagenome, the script `run_metabat2.sh` will be used along with a list of the mixed-community metagenome sample names in a file `metagenome_samples.txt`. It is assumed that reads were mapped to the filtered scaffolds that are to be binned, and the corresponding `.bam` files are in the folder `map_files_filt`.

```
cat metagenome_samples.txt

ERR598984
ERR599005
ERR599006
ERR599017
ERR599044
ERR599126
ERR599133
ERR599148
ERR599165
ERR599174
```

The mapping files are not required to run metaBAT2, but they will substantially increase its accuracy and number of bins generated.

```
#!/bin/sh
# run_metabat2.sh

THREADS="16"
SAMPLES="metagenome_samples.txt"

mkdir -p metabat2_bins && cd metabat2_bins

while IFS= read -r SAMPLE || [ -n "$SAMPLE" ]
do
    # Skip empty lines
    if [ -z "$SAMPLE" ]; then
        continue
    fi

    echo "Binning sample $SAMPLE"

    # Command to run for each sample
    runMetaBat.sh \
        "../filtered_scaffolds/${SAMPLE}_scaffolds.fasta" \
        "../map_files_filt/${SAMPLE}.bam"
```

```
        --minContig 2500 \
        --maxP 95 \
        --minS 60 \
        --minCV 1 \
        --minClsSize 200000 \
        --numThreads "$THREADS"
    echo "Done ${SAMPLE}"
done < ../"$SAMPLES"
```

Here, the flags `–minContig 2500`, `–maxP 95`, `–minS 60`, `–minCV 1`, `–minClsSize 200000` are the default parameters for metaBAT2, but the user should consider each when running metaBAT2. They each can affect the number of bins generated and the overall quality of the bins, with a trade-off between maximizing recovery and increasing contamination. Assessing the quality of bins and understanding contamination will be addressed below.

The script will create a folder named `metabat2_bins` that will contain one subfolder per sample, each containing the MAGs binned for that sample. Across all 10 mixed-community metagenome samples, a total of 436 MAGs were generated, ranging from as few as 23 to as many as 86 in any given sample (Figure 2A). The number of MAGs binned by metaBAT2 increased with the number of scaffolds, total number of bases, and the N50 of the input assembly (Figure 2B).

An individual MAG roughly represents a microbial population at the species or genus level (2, 66, 67). That does not necessarily mean that each MAG represents a complete genome. In fact, the completeness and contamination (sequences originating from another organism) for MAGs can be estimated quite confidently with MAG QC. It is generally accepted that a MAG with at least 90% estimated completeness and at most 10% estimated contamination is a ‘high-quality’ MAG (26, 68). High-quality meaning the MAG is an accurate and near-complete representation of a microbial population (26). MAGs with at least 50% completeness and at most 10% contamination are of medium-quality. They are still useful and adequate for population and community level analyses (68), but interpretations based on them should be formed and considered with caution. Any MAG that both has an estimated completeness less than 50% and an estimated contamination over 10% is of low-quality (68) and should not be further analyzed with the remaining medium- and high-quality MAGs.

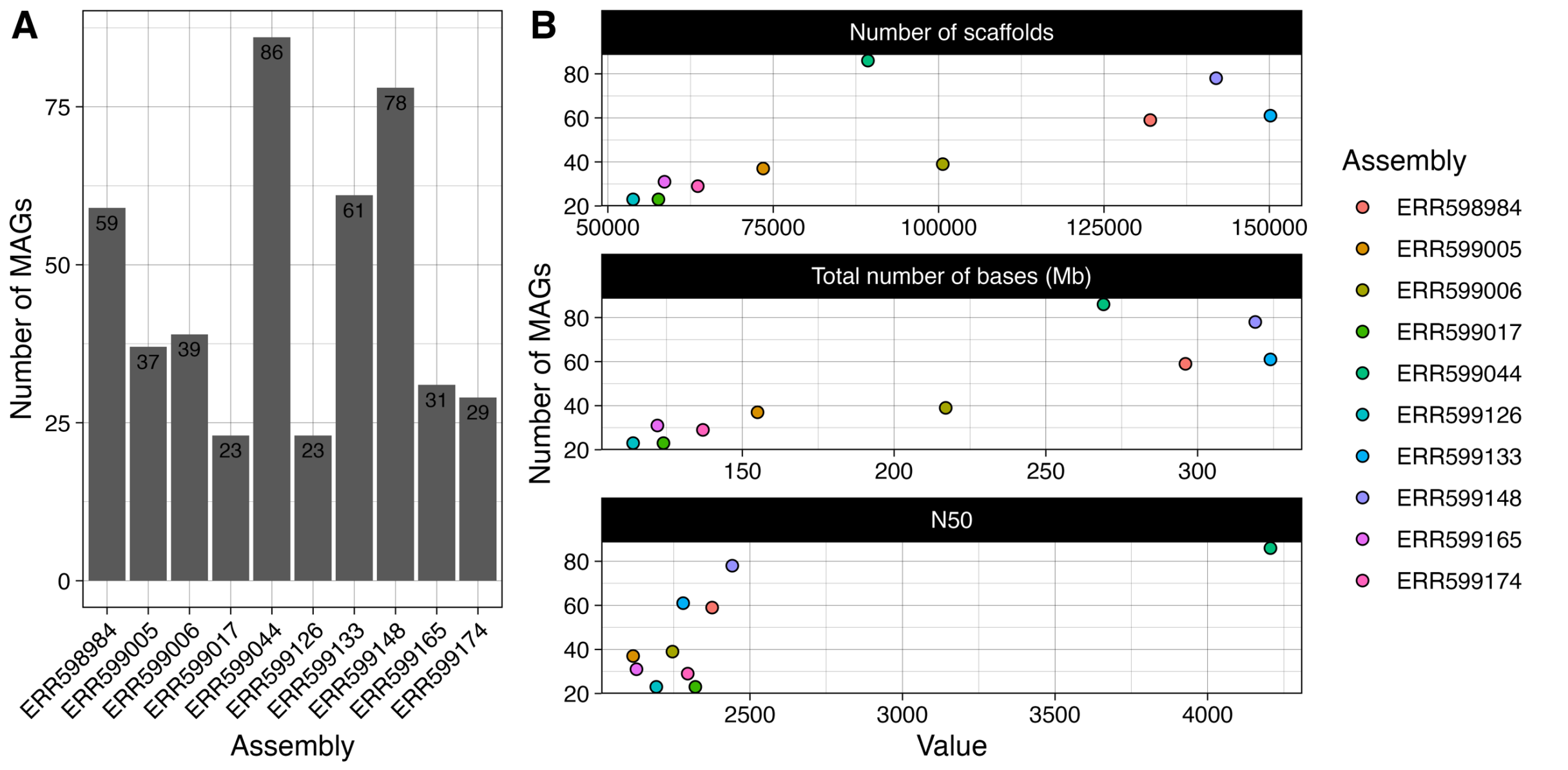

**Figure 2: MAGs Binned Per Assembly. (A)** Number of MAGs generated for each *Tara* mixed-community metagenome assembly. **(B)** Number of MAG versus total number of scaffolds in the input assembly, total number of bases in the input assembly, and input assembly N50. Statistics calculated for scaffolds > 1,000 bp as demonstrated in Section 3.5.

After metaBAT2 has finished binning each mixed-community metagenome sample, the quality of each will be assessed with the tool CheckM (26) using the script `run_checkm.sh`:

```
#!/bin/sh
# run_checkm.sh

THREADS="16"
SAMPLES="metagenome_samples.txt"

mkdir -p checkm

while IFS= read -r SAMPLE || [ -n "$SAMPLE" ]
do
    # Skip empty lines
    if [ -z "$SAMPLE" ]; then
        continue
    fi

    # Find the path to the metaBAT2 folder for the current sample
    # Using 'find' with wildcards '*' because folder names can vary
    # Adjust if your metaBAT2 outputs have a different pattern
    MAG_DIR=$(find metabat2_bins -name ${SAMPLE}_*.metabat-bins-*)

    echo "Running CheckM on sample $SAMPLE"
```

```
    # Command to run for each sample
    checkm lineage_wf \
        --extension fa \
        --file "checkm/${SAMPLE}/${SAMPLE}_quality.txt" \
        --tab_table \
        --threads "$THREADS" \
        "$MAG_DIR" \
        "checkm/${SAMPLE}"
    echo "Done ${SAMPLE}"
done < "$SAMPLES"
```

This script will use the sample names in `metagenome_samples.txt` again to run checkM on each output folder created by metaBAT2. The folder named are inferred using the find command since the `runMetaBat.sh` command invoked by the `run_metabat2.sh` script above will create variable folder names based on the dates and times of execution. Here, the lineage workflow of checkM is executed with `lineage_wf`. This workflow obtains completeness and contamination estimates using different sets of marker genes for each 'lineage' the query MAGs were placed in by checkM. MetaBAT2 writes the output MAGs to files ending in `.fa`, which is different than the input file extension `.fna` that checkM assumes. This extension is provided to checkM with the `–extension` flag. The flags `–file` and `–tab_table` will write the results to a tab-delimited table that can be read later. As usual, `–threads` sets the maximum number of CPUs to use.

The checkM output files should be in the new folder `checkm` with one subfolder per sample. For each set of input MAGs per sample, their quality estimations are found in a table with a name ending in `_quality.txt`. These can each be manually inspected to sort high, medium, and low-quality MAGs. Alternatively, one can use a language like Python and write a script to automatically do this. Such a script `sort_mags.py` is provided below:

```
# sort_mags.py

import os
import csv
from pathlib import Path

def process_sample(sample, checkm_path, metabat2_bins_path,
                   sorted_mags_path, summary_writer):
    quality_file = checkm_path / f"{sample}/{sample}_quality.txt"

    # Check if the quality file exists before proceeding
    if not quality_file.exists():
        print(f"Quality file not found for sample {sample}, skipping...")
        return

    bins_dir = next(metabat2_bins_path.glob(f"{sample}_*.metabat-bins-*"),
                    None)
```

```
    if not bins_dir:
        print(f"Bins directory not found for sample {sample}, skipping...")
        return

    with quality_file.open() as f:
        reader = csv.DictReader(f, delimiter='\t')
        for row in reader:
            bin_id = row['Bin Id']
            new_bin_id = f"{sample}__{bin_id.replace('bin.', 'bin_')}"
            completeness = float(row['Completeness'])
            contamination = float(row['Contamination'])

            if completeness >= 90.0 and contamination <= 10.0:
                quality = "high_quality"
            elif 50.0 <= completeness < 90.0 and contamination <= 10.0:
                quality = "medium_quality"
            else:
                quality = "low_quality"

            fa_file_path = bins_dir / f"{bin_id}.fa"
            if fa_file_path.exists():
                quality_dir = sorted_mags_path / quality
                quality_dir.mkdir(parents=True, exist_ok=True)
                symlink_filename = f"{new_bin_id}.fa"
                symlink_path = quality_dir / symlink_filename
                if not symlink_path.exists():
                    symlink_path.symlink_to(fa_file_path.resolve())

                summary_writer.writerow([bin_id, sample,
                                         new_bin_id, completeness,
                                         contamination, quality])

def main(checkm_path_str, metabat2_bins_path_str, sorted_mags_path_str):
    checkm_path = Path(checkm_path_str)
    metabat2_bins_path = Path(metabat2_bins_path_str)
    sorted_mags_path = Path(sorted_mags_path_str)

    # Create sorted_mags directory and its subdirectories
    sorted_mags_path.mkdir(parents=True, exist_ok=True)

    summary_file_path = sorted_mags_path / "checkm_summary.txt"
    with summary_file_path.open('w', newline='') as summary_file:
        writer = csv.writer(summary_file, delimiter='\t')
        writer.writerow(["BinID", "Sample", "NewBinID",
                         "Completeness", "Contamination",
                         "Quality"])

        for sample_dir in checkm_path.iterdir():
            if sample_dir.is_dir():
```

```
                sample = sample_dir.name
                process_sample(sample, checkm_path,
                               metabat2_bins_path,
                               sorted_mags_path, writer)

if __name__ == "__main__":
    checkm_path = "checkm"
    metabat2_bins_path = "metabat2_bins"
    sorted_mags_path = "sorted_mags"
    main(checkm_path, metabat2_bins_path, sorted_mags_path)
```

Assuming the user has Python (versions 3 and above) installed, the script can be run with `python3 sort_mags.py`. This script will create links to the original `.fa` files produced by metaBAT2 (so don't remove the originals, they are not copies, just shortcuts!). The links will be created beneath the new folder `sorted_mags` be sorted into sub folders `high_quality`, `medium_quality`, and `low_quality` based on the thresholds described above, but they can be adjusted. A table named `checkm_summary.txt` will be written by this script that will display the determined qualities for each bin. The filenames for the links will be prepended with their sample name, so `checkm_summary.txt` also provides a key to convert the old MAG names to the new ones:

```
head sorted_mags/checkm_summary.txt

BinID   Sample    NewBinID            Completeness Contamination   Quality
bin.1   ERR599006 ERR599006__bin_1       46.19         1.72      low_quality
bin.10  ERR599006 ERR599006__bin_10      43.26         3.45      low_quality
bin.11  ERR599006 ERR599006__bin_11      40.52         0.0       low_quality
bin.12  ERR599006 ERR599006__bin_12      13.79         0.0       low_quality
bin.13  ERR599006 ERR599006__bin_13      91.87         1.08      high_quality
bin.14  ERR599006 ERR599006__bin_14      6.03          0.0       low_quality
bin.15  ERR599006 ERR599006__bin_15      0.0           0.0       low_quality
bin.16  ERR599006 ERR599006__bin_16      53.85         0.43      medium_quality
bin.17  ERR599006 ERR599006__bin_17      43.97         10.34     low_quality
```

Using this table, the number of low, medium, and high-quality MAGs can be counted using `grep`:

```
grep -c "low_quality" sorted_mags/checkm_summary.txt
grep -c "medium_quality" sorted_mags/checkm_summary.txt
grep -c "high_quality" sorted_mags/checkm_summary.txt

324
90
22
```

The total number of medium- and high-quality MAGs is 90+22 = 112. For 10 complex, mixed-community metagenome samples, that is a good amount to work with for downstream ecological and functional analyses. These 112 MAGs will be used in the further sections. If one wishes to attempt to get high-quality MAGs out of the low-quality ones, *bin refinement* is worth consideration. Bin refinement describes the manual or automated process of assigning and re-assigning scaffolds to MAGs based on sequence composition, GC skew, and gene annotations (69, 70). Automated bin refinement can be performed using the tool metaWRAP (15). Manual bin refinement can be achieved in a visual and interactive interface within the suite of (meta)genomics tools, Anvi'o, (18).

The scaffolds from the 10 mixed-community samples that were either unbinned or ended up in low-quality bins are still useful. They contain potentially interesting gene-level information that can be leveraged later, however they should not be used for genome-based analyses. The links to the medium- and high-quality mags can be copied (preserving as links to save on storage) and placed into their own combined folder `sorted_mags/medium_high_quality` for further analysis.

```
mkdir orted_mags/medium_high_quality
cp -P sorted_mags/medium_quality/*.fa sorted_mags/high_quality/*.fa sorted_ma
gs/medium_high_quality
```

### *3.7 - Assigning Taxonomy to Microbial Bins*

At this point, the mixed-community metagenomes have been assembled, QC'd, binned into MAGs, and filtered even further to determine which MAGs represent medium- or high-quality genomes. It has not yet been determined just which organisms these MAGs represent. The taxonomy of these MAGs may be informative on their functional potential and possible evolutionary histories. For whole genome sequences, it is typical to obtain taxonomic assignments using a 16S rRNA marker gene approach (71, 72), a multi-gene tree approach (73, 74), a whole-genome distance-based approach (75, 76), or a combination of these approaches (77). These approaches are useful for isolate genomes, but they tend to perform poorly when applied to MAGs. Since MAGs tend to be incomplete, then there may not be 16S rRNA genes or other marker genes on the genome. Also, since MAGs can vary greatly in length, completeness, and contamination, methods that leverage whole-genome distances will not always work. There is a robust, highly standardized tool that is specifically designed to obtain taxonomic assignments for MAGs. The tool GTDB-Tk (27, 78) can take fragmented, potentially contaminated genomes and assign taxonomy using a comprehensive database of isolate and uncultivated genomes (79) in a high-throughput way.

Here, GTDB-Tk will be run using the `classify_wf` option, which will obtain taxonomic assignments for the input medium- and high-quality MAGs by placing them in the GTDB reference phylogenomic trees. Since all the MAGs to be analyzed are in the folder `sorted_mags/medium_high_quality`, this directory can be given to the `gtdbtk classify_wf` command. The extension of the MAG files will be specified (`.fa`) and the taxonomy results will be written inside the folder `gtdb_taxonomy`:

```
gtdbtk classify_wf \
    --genome_dir sorted_mags/medium_high_quality \
    --out_dir gtdb_taxonomy \
    --extension fa \
    --cpus 16
```

The results will be written to `gtdb_taxonomy/gtdbtk.ar53.summary.tsv` for identified Archaea and `gtdb_taxonomy/gtdbtk.bac120.summary.tsv` for identified Bacteria.

### *3.8 - Identifying Viruses and Other Mobile Genetic Elements*

It is estimated that there are $10^{31}$ viruses on earth, primarily viruses of microbes (80). As predators of microbes, viruses have great ecological impacts on their host populations (81, 82), and, by extension, biogeochemical impacts on their surrounding environments (83–85). In addition to the 10 mixed-community metagenomes that were used to bin microbial genomes into MAGs, the *Tara* oceans data used here contains 10 viromes that where generated to enrich for viral DNA. However, it should not be assumed that all scaffolds among the viromes are viral sequences. Although viromes generally outperform mixed-community metagenomes in maximizing observable virus abundance and diversity (37), non-viral DNA can still end up in viromes after enrichment (40, 58). Additionally, when sequencing metagenomes of whole microbial communities, one can expect to sequence viral DNA either from viruses in an extracellular state or from viruses integrated in their host genomes. Because of the imperfect enrichment for viruses in viromes and the presence of viruses among microbes, viral scaffolds need to be identified and sorted from non-viral scaffolds in both viromes and mixed-community metagenomes.

Several tools have been developed in recent years to identify viral sequences in metagenome data, but the most robust, accurate, and user-friendly ones to date are geNomad (28), VIBRANT (86), and VirSorter2 (87). Each tool performs similar when benchmarked against each other (28, 86), but each have their own unique qualities that may influence the researcher's decision to use one over the other. For example, VirSorter2 contains different classifiers that are specialized to specific types of viruses (87). VIBRANT produces additional outputs for the identification of putative *auxiliary metabolic genes* (86). Additionally, geNomad identifies plasmids as well as viruses, and it attempts to obtain taxonomic assignments to identified viruses (28). Since plasmids may also be of interest to the researcher in a metagenomic study in addition to viruses, geNomad will be used to identify viral and plasmid scaffolds in all 20 *Tara* assemblies here. When running, geNomad is implemented as a series of modules. They can all be executed sequentially by executing `genomad end-to-end`. This will be done for every assembly of filtered scaffolds using the script `run_genomad.sh`:

```
#!/bin/sh
# run_genomad.sh

SCAFFOLD_DIR="filtered_scaffolds"
OUTPUT_DIR="genomad_out"
DB_DIR="/path/to/genomad/db"
THREADS="16"

mkdir -p "$OUTPUT_DIR"

for SCAFF in ${SCAFFOLD_DIR}/*fasta; do
    # Extract the sample name
    SAMPLE=$(basename "$SCAFF" _scaffolds.fasta)

    # Run geNomad
    genomad end-to-end \
```

```
        --cleanup \
        --threads "$THREADS" \
        "$SCAFF" \
        "${OUTPUT_DIR}/${SAMPLE}" \
        "$DB_DIR"

    echo "Done ${SAMPLE}"
done
```

The `DB_DIR` variable will need to be set to the full path to the local copy of the geNomad database created during installation. The `-cleanup` flag removes intermediate files after geNomad finishes a run on a sample, to save on storage. The results for will be located beneath the folder `genomad_out` with subfolders for each sample. The final virus and plasmid results will be in another subfolder with the suffix `_summary`. For example, the folder `genomad_out/ERR599174/ERR599174_scaffolds_summary` contains virus and plasmid results for sample ERR599174:

```
ls genomad_out/ERR599174/ERR599174_scaffolds_summary

ERR599174_scaffolds_plasmid.fna
ERR599174_scaffolds_plasmid_genes.tsv
ERR599174_scaffolds_plasmid_proteins.faa
ERR599174_scaffolds_plasmid_summary.tsv
ERR599174_scaffolds_virus.fna
ERR599174_scaffolds_virus_genes.tsv
ERR599174_scaffolds_virus_proteins.faa
ERR599174_scaffolds_virus_summary.tsv
```

For this sample, all identified virus nucleotide sequences are in the file `ERR599174_scaffolds_virus.fna` and all identified plasmid nucleotide sequences are in `ERR599174_scaffolds_plasmid.fna`. Additionally, the files ending with `_summary.tsv` contain scaffold-level information for the identified viruses and plasmids.

To do a quick check of how many viruses and plasmids were identified in this sample, `grep` can be used:

```
grep -c ">" genomad_out/ERR599174/ERR599174_scaffolds_summary/ERR599174_scaff
olds_virus.fna
grep -c ">" genomad_out/ERR599174/ERR599174_scaffolds_summary/ERR599174_scaff
olds_plasmid.fna

1170
151
```

A substantial number of viral scaffolds were identified in this sample. While the plasmid sequences identified by geNomad may be of interest to the researcher, but they will not be further investigated here.

Combining all `virus_summary.tsv` files can allow for the number of viral scaffolds across all samples to be visualized. Shown in Figure 3 are the fraction and total number of scaffolds identified as viral for the mixed-community metagenome and virome assemblies of each *Tara* sample. As expected, the viromes in each assembly pair yielded a greater fraction of viral scaffolds and a greater total number of viral scaffolds.

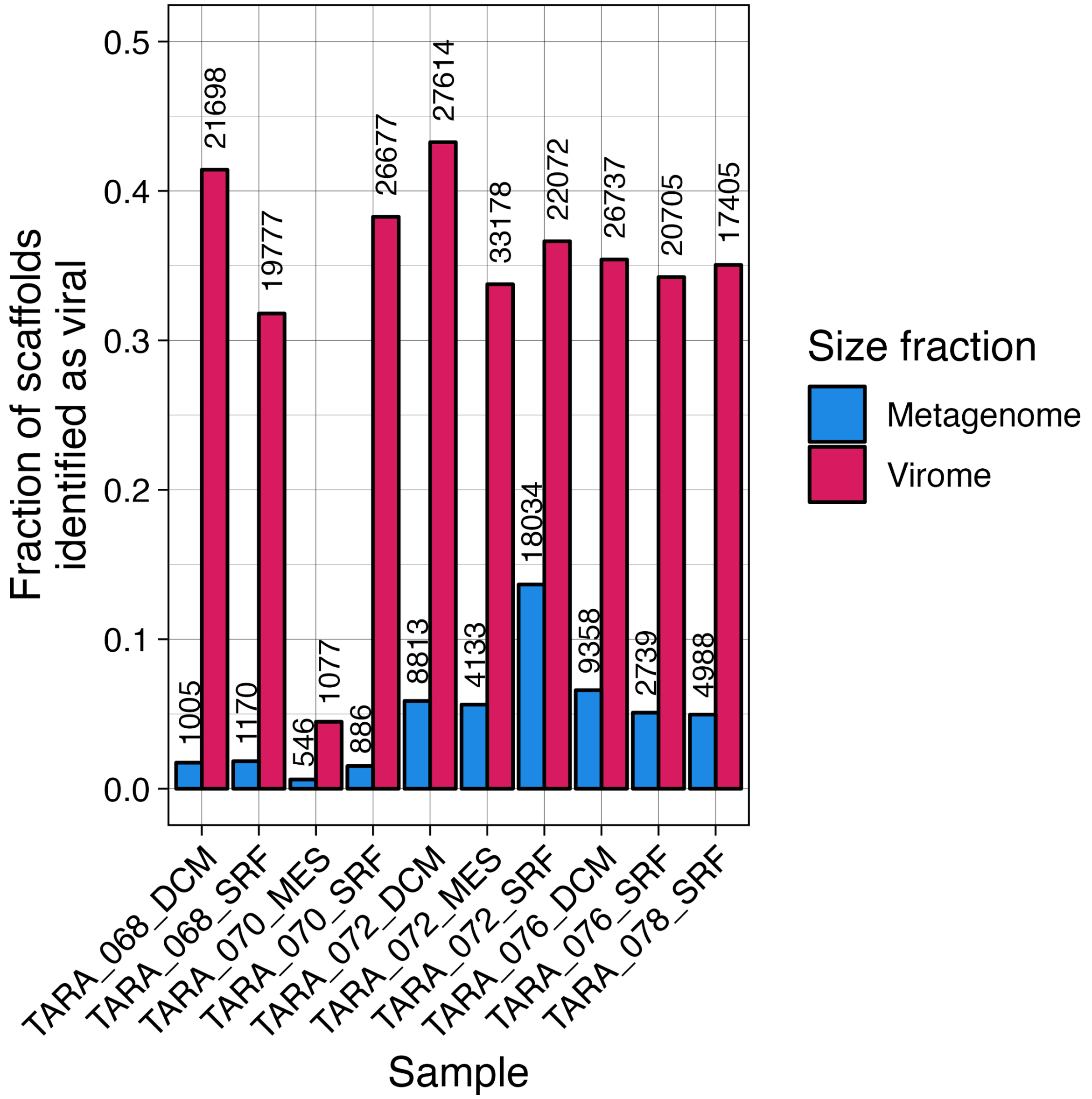

**Figure 3: Viral Scaffolds Identified Per Sample.** Viral scaffolds identified by geNomad for each *Tara* oceans sample as a fraction of all scaffolds > 1,000 bp in the input assembly are given. Bar colors correspond to the virome and mixed-community metagenome size fraction assemblies for each Tara sample. Atop each bar is the total number of viral scaffolds identified in the corresponding assembly.

### *3.9 - Binning Viral Genomes*

In Section 3.6, scaffolds were binned into MAGs. This was because microbial genomes appear highly fragmented in metagenomic data, and reconstructing genomes into MAGs can improve genomic and population level analyses. It is also common for viral genomes to be fragmented in mixed-community metagenomes and viromes. Although viruses typically have genome sizes orders of magnitude smaller than their microbial hosts, genome fragmentation can make metagenomic analyses of viral communities a challenge (30, 88). Viral genome fragments can be binned into *viral metagenome-assembled genomes* (vMAGs) to overcome some of these challenges.

Binning viral genomes is not (yet) as common as binning microbial genomes into MAGs, primarily due to the small size of viral genomes and the high complexity and mosaicism of viral scaffolds (89). However, there are several benefits to binning viral scaffolds into vMAGs. First, vMAGs can accurately reconstruct more complete viral genomes than individual scaffolds (29, 90, 91). Second, binning scaffolds of the same viruses into vMAGs can lead to a more accurate representation of viral diversity in a community, since treating single scaffolds as an individual genome may overestimate diversity (92). Third, there are some viruses with large genomes such as 'jumbo', 'huge', and 'mega' phage (93), and NCLDVs (a.k.a. 'giant viruses') which infect microbial eukaryotes (94). Binning viral scaffolds may improve one's ability to detect these unique viruses in metagenomic data (95).

There are some drawbacks to binning viral scaffolds into vMAGs. Most viral metagenomic tools are not designed to handle vMAGs, rather, they assume individual scaffolds represent a single viral genome. This requires further processing and manipulation of vMAGs so they can be compatible with such tools (see Section 3.10, Section 3.11, Section 3.12). Also, just as with binning microbial genomes, binning viral genomes can produce contaminated bins. However, vMAGs can be QC'd like MAGs, both with computational tools and with manual curation.

Although there are many tools available designed to bin microbial scaffolds (see Section 3.6), these tools typically do not perform well when binning viral scaffolds (29, 90), except for the kinds of very large viruses mentioned above. This is primarily because viruses do not have universal marker genes like bacteria and archaea, which several microbial binning tools search for in the binning process. Although binning viral scaffolds into vMAGs is mechanistically difficult, there are a handful of tools available specifically designed to generate vMAGs. vRhyme bins viral scaffolds using single- or multi-sample read coverage and sequence features (29). COBRA bins viruses by resolving metagenome assembly breakpoints in assembly graphs (90). ViralCC uses metagenomic Hi-C data for genome binning (92). And, although it is not a tool, PHAMB is a framework for binning viruses using random forest models (91).

Here, vRhyme will be used to bin viral scaffolds into vMAGs since its input requirements (viral scaffolds and read mapping data) were already generated. Although coassembly was not performed here, it is worth noting that vRhyme is particularly useful for coassemblies or other metagenomic datasets with biological replicates. This is because of its unique ability to use multi-sample read coverage metrics in binning (29). Mapping reads from replicated samples can greatly improve vRhyme's accuracy and recall (29), the researcher is encouraged to consider this in their own workflow. The script `run_vrhyme.sh`, shown below, will be used to execute vRhyme for every set of viral scaffolds identified by geNomad.

```
#!/bin/sh
# run_vrhyme.sh

THREADS="16"
INDIR="genomad_out"
OUTDIR="vrhyme_bins"
BAMDIR="map_files_filt"


mkdir -p $OUTDIR && cd $OUTDIR

for DIR in ../$INDIR/ERR*; do
    SAMPLE="${DIR##*/}"
    echo "Binning sample: $SAMPLE"

    # vRhyme command to run for each assembly
    vRhyme \
        -i "../${INDIR}/${SAMPLE}/${SAMPLE}_scaffolds_summary/${SAMPLE}_scaff
olds_virus.fna" \
        -b "../${BAMDIR}/${SAMPLE}.bam" \
        -o "../${OUTDIR}/${SAMPLE}" \
        -l 2000 \
        --prefix "$SAMPLE"__

    # Add assembly prefix to vRhyme bin files
    for VMAG in ../${OUTDIR}/${SAMPLE}/vRhyme_best_bins_fasta/*.f*; do
        BASE=$(basename ${VMAG})
        mv -- "$VMAG" "../${OUTDIR}/${SAMPLE}/vRhyme_best_bins_fasta/${SAMPLE
}__${BASE}"
    done

    # Extract the unbinned viral sequences with a vRhyme auxiliary script
    extract_unbinned_sequences.py \
        -i ../${OUTDIR}/${SAMPLE}/vRhyme_best_bins.*.membership.tsv \
        -f "../${INDIR}/${SAMPLE}/${SAMPLE}_scaffolds_summary/${SAMPLE}_scaff
olds_virus.fna" \
        -o "../${OUTDIR}/${SAMPLE}/${SAMPLE}_unbinned_viruses.tmp.fna"

    # Add assembly prefix to headers of unbinned sequences
    UNBINNED_TMP="../${OUTDIR}/${SAMPLE}/${SAMPLE}_unbinned_viruses.tmp.fna"
    UNBINNED="../${OUTDIR}/${SAMPLE}/${SAMPLE}_unbinned_viruses.fna"
    awk -v PREFIX="$SAMPLE" '/^>/{print ">" PREFIX "__" substr($0, 2); next}
{print}' \
        "$UNBINNED_TMP" \
```

```
        >> "$UNBINNED"
    rm "$UNBINNED_TMP"

    # Split the unbinned sequences into individual fasta files
    UNBINNED_DIR="../${OUTDIR}/${SAMPLE}/vRhyme_unbinned_viruses_fasta"
    mkdir -p "$UNBINNED_DIR"
    if [ -f "$UNBINNED" ]; then
        COUNTER=1
        awk -v SAMPLE="$SAMPLE" -v OUTDIR="$UNBINNED_DIR" \
            '/^>/ {if(x>0) close(OUTNAME); \
            OUTNAME=sprintf("%s/%s__vRhyme_unbinned_%d.fasta",OUTDIR,SAMPLE,+
+x);} \
            {print > OUTNAME}' "$UNBINNED"
    fi

    # 'N-link' bins for CheckV with a vRhyme auxiliary script
    link_bin_sequences.py \
        -i "../${OUTDIR}/${SAMPLE}/vRhyme_best_bins_fasta" \
        -o "../${OUTDIR}/${SAMPLE}/vRhyme_best_bins_fasta_linked" \

    # Add assembly prefix to headers of linked bin sequences
    find "../${OUTDIR}/${SAMPLE}" \
        -type f \
        -path "*/vRhyme_best_bins_fasta_linked/*.fasta" | while read -r FASTA
; do
        TMP=$(mktemp) # Create a temporary file
        awk -v SAMPLE="$SAMPLE" '/^>/ {$0=">" SAMPLE "__" substr($0, 2)} 1' \
            "$FASTA" > "$TMP"
        mv "$TMP" "$FASTA"
    done

    echo "Done ${SAMPLE}"
done
```

There are multiple components to this script:

1. An output folder `vrhyme_bins` is created with one subfolder per metagenome/virome assembly. vRhyme is executed to bin all viral scaffolds identified by geNomad, for each assembly. The flag `-i` provides the input viral scaffolds, `-b` provides the mapped read `.bam` files previously generated that are required by vRhyme, `-o` sets the output directory for that assembly, `-l 2000` sets the input scaffold length requirement to 2,000 bp as recommended by vRhyme, and `–prefix “$SAMPLE”__` adds the assembly ID as a prefix to the output bin sequence names (with an added double-underscore to easily distinguish the assembly IDs from the scaffold names).
2. The same assembly ID prefix is added to the output bin filenames, since vMAGs from all assemblies will be combined and used below.
3. The viral scaffolds that were not binned by vRhyme are extracted using the vRhyme auxiliary script `extract_unbinned_sequences.py`. They are split up into individual files with assembly ID prefixes in their names and placed into a folder named `vRhyme_unbinned_viruses_fasta`.

4. Scaffolds in each bin are joined by a string of Ns into a single sequence using the vRhyme auxiliary script `link_bin_sequences.py` and are placed into a separate folder `vRhyme_best_bins_fasta_linked`. This is to QC the bins using a tool that expects a single scaffold per virus, below. Assembly ID prefixes are added to their sequence headers, too.

For most samples, around half of all input viral scaffolds at least 2,000 bp were binned (Figure 4a), with an average of around 5 scaffolds per bin (Figure 4b). This yields over 156,000 vMAGs across all samples and assemblies (here, both binned and unbinned viral genomes are referred to as vMAGs).

The completeness and contamination of each vMAG can be estimated with the tool CheckV (30). CheckV estimates completeness and contamination with a combined marker gene, amino-acid identity, and sequence feature approach (30). CheckV was designed to handle single-scaffold viral genomes rather than vMAGs. To run CheckV, the N-linked vMAGs generated by the `run_vrhyme.sh` script above will be combined with the single-scaffold vMAGs with the following command:

```
find vrhyme_bins/ \
    -type f \( \( \
    -path "*/vRhyme_best_bins_fasta_linked/*.fasta" \) \
    -o \( -name "*_unbinned_viruses.fna" \) \) \
    -print0 \
    | xargs -0 cat > virus_genomes_linked.fna
```

Now, the quality of every vMAG can be estimated with CheckV:

```
checkv end_to_end \
    virus_genomes_linked.fna \
    checkv_out \
    -t 16
```

The CheckV results will be deposited in a folder `checkv_out`. The final quality estimates for every vMAG are given in the table `quality_summary.tsv`. If one wants to take a conservative approach when analyzing viral genomes, just the medium-quality, high-quality, and/or complete vMAGs can be retained for further analysis. It is not always necessary to filter low-quality MAGs from analyses. For instance, since there is no universal marker gene for viruses, it is not possible to assess the diversity of viral communities with an approach like 16S amplicon sequencing. To obtain accurate estimates of viral diversity or community composition, it is best to retain all identified viral scaffolds (if they are dereplicated and at least 10 Kb, see Section 3.10). Additionally, CheckV's marker-based completion estimates are biased towards well-studied and characterized viruses. This is no fault of CheckV itself, which is why CheckV assigns confidence levels to each assignment. This means, when analyzing understudied environments, some 'low-quality' designations may be false negatives that are the result of unidentifiable marker genes in rare and highly novel viruses (30). For these reasons, the vMAGs will not be sorted and filtered by CheckV quality like the MAGs were using CheckM quality.

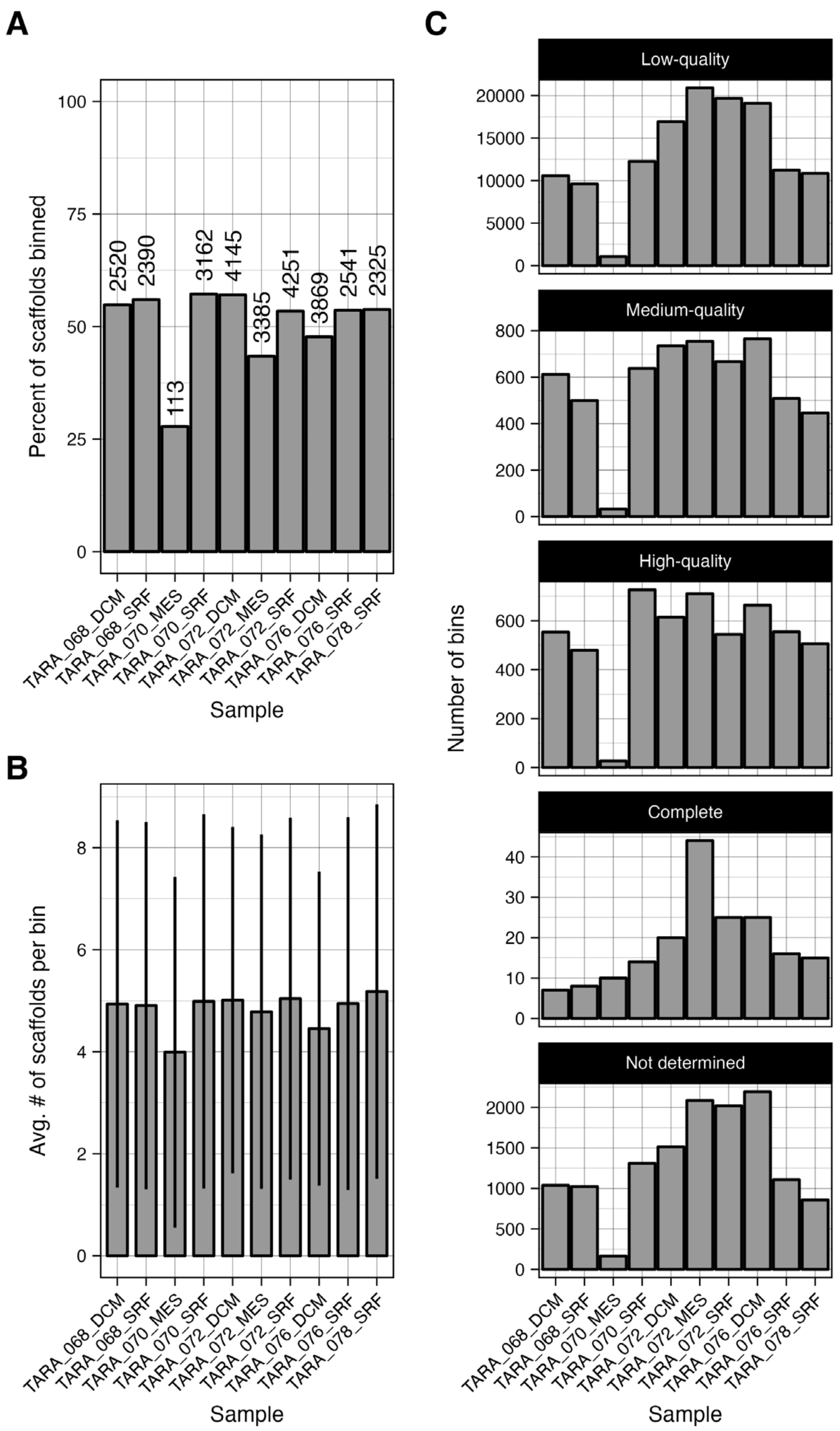

**Figure 4: vRhyme Binning and vMAG Quality Summary. (A)** Percent of viral scaffolds in each *Tara* sample that were binned. Values atop each bar give the total number of multi-scaffold bins in that sample. **(B)** Average number of scaffolds assigned to each bin for every *Tara* sample. Whiskers represent standard deviations. **(C)** Total number of vMAGs (binned and unbinned) assigned to each CheckV quality for every *Tara* sample.

### *3.10 - Microbial and Viral Genome Dereplication*

In metagenomics/viromics, genome dereplication is the process of identifying and consolidating redundant genomes that represent the same (or very similar) microbial/viral species within a dataset (31, 96). This is crucial in studies with multiple metagenomes and/or viromes because it can greatly reduce computational redundancy. Dereplication also improves the clarity of downstream abundance-based microbial and viral diversity assessments since it ensures that each unique genomic entity is represented once. Also, by selecting the best representatives among genome species clusters, dereplication ensures that subsequent analyses are based on the best-available data. However, a potential downside is the risk of oversimplification. The removal of closely related genomes could inadvertently eliminate valuable information about fine-scale genetic diversity and evolutionary dynamics within microbial and viral populations (96). On the other hand, strain-level variation can still be analyzed with dereplicated MAGs/vMAGs with read-mapping based diversity calculations (67, 97). Overall, dereplication is highly recommended for analyses of multiple microbial and viral communities.

Tools like dRep (31) have been specifically designed for MAG dereplication and selection of the ‘best’ genomes. Although dRep was primarily designed for microbial genomes, it can be used with viruses/vMAGs by altering some parameters. The vMAGs will be dereplicated first. Before dRep is executed, a list of the vMAG file paths need to be created so dRep can identify and read the sequences within each individual vMAG. This can be done with the script `make_genome_list_virus.sh`, below:

```
#!/bin/sh
# make_genome_list_virus.sh

BASE_DIR="vrhyme_bins"
OUTPUT_FILE="virus_genome_paths.txt"

# Clear the output file to start fresh
> "$OUTPUT_FILE"

# Find and add paths from vRhyme_best_bins_fasta folders
find "$BASE_DIR" \
    -type f \
    -path "*/vRhyme_best_bins_fasta/*.fasta" \
    >> "$OUTPUT_FILE"

# Find and add paths from vRhyme_unbinned_viruses_fasta folders
find "$BASE_DIR" \
    -type f \
    -path "*/vRhyme_unbinned_viruses_fasta/*.fasta" \
    >> "$OUTPUT_FILE"
```

The script will create the file `virus_genome_paths.txt`. This file can be given to the `dRep dereplicate` command:

```
dRep dereplicate \
    --processors 16 \
    --genomes virus_genome_paths.txt \
    --length 10000 \
    --S_algorithm skani \
    --ignoreGenomeQuality \
    -pa 0.8 -sa 0.95 -nc 0.85 \
    -comW 0 -conW 0 -strW 0 \
    -N50W 0 -sizeW 1 -centW 0 \
    --skip_plots \
    dereplicated_viruses
```

By default, dRep forms species-level clusters at an average nucleotide identity of 95%. It considers MAG completeness, contamination, heterogeneity, N50, centrality, and genome size for selecting the best MAG representative of a species-level cluster. Except for genome size, these metrics are not optimal for dereplicating viral genomes (see 'Comparing and dereplicating non-bacterial genomes' from the dRep manual (31)), so they need not be considered. To achieve this, the flags `–ignoreGenomeQuality -comW 0 -conW 0 -strW 0 -N50W 0 -sizeW 1` and `-centW 0` are added. This will effectively make the chosen representative in each species cluster the largest vMAG. Since over 156,000 vMAGs from all samples and assemblies will be dereplicated at once, the flags `–S_algorithm skani` and `–skip_plots` are added to improve computational efficiency and decrease runtime. The flags `-pa 0.8 -sa 0.95 -nc 0.85` set the primary and secondary cluster ANI cutoffs to 80% and 95%, respectively. A primary cluster ANI cutoff of 80% is recommended because of the high variability of viral genomes, and the secondary cluster ANI cutoff of 95% will result in (roughly) viral species-level clusters (43, 98). The flag `-nc 0.85` sets a genome overlap of 85% when comparing genomes.

Importantly, the flag `–length 10000` is included. This sets a minimum length requirement of 10 Kb for input genomes, which means any vMAG less than 10 Kb will not end up in the set of dereplicated vMAGs. This minimum can be set to whatever the user desires, however, a minimum genome length of 10 Kb has been demonstrated to be an optimal requirement for viral community analyses (43). Unless a significant fraction of the researcher's vMAGs are below this cutoff, or the study in question is focused on viruses with ultra-small genomes, we recommend using this 10 Kb minimum.

The dRep output will be in the `folder dereplicated_viruses`. Beneath, the final set of vMAG species representatives are found in the folder `dereplicated_genomes`. These dereplicated vMAGs can be combined into one single file for use later:

```
cat dereplicated_viruses/dereplicated_genomes/*.fasta \
        > virus_genomes_dereplicated.fna
```

For the microbial genomes, a list of the paths to the medium- and high-quality MAGs can be created using a similar script as used for the vMAGs, `make_genome_list_mags.sh`:

```
#!/bin/sh
# make_genome_list_mags.sh
```

```
BASE_DIR="sorted_mags/medium_high_quality"
OUTPUT_FILE="mag_genome_paths.txt"

# Clear the output file to start fresh
> "$OUTPUT_FILE"

# Find and add paths from medium and high quality MAGs
find "$BASE_DIR" \
    -type l \
    -name "*.fa" \
    >> "$OUTPUT_FILE"
```

The file `mag_genome_paths.txt` can be given to the `dRep dereplicate` command. Since MAGs are being dereplicated this time instead of vMAGs, most of the default dRep options are used:

```
dRep dereplicate \
    --processors 16 \
    --genomes mag_genome_paths.txt \
    --skip_plots \
    dereplicated_mags
```

The microbial dRep output will be in the folder `dereplicated_mags`. Beneath, the final set of MAG species representatives are found in the folder `dereplicated_genomes`. Assembly ID prefixes can be added to the scaffold names in these MAGs using the script `mag_header_prefix.sh`, below, so all MAGs be combined into one file and used later as needed.

```
#!/bin/sh
# mag_header_prefix.sh

EXT=".fa"
INDIR="dereplicated_mags/dereplicated_genomes"
OUTDIR="dereplicated_mags/dereplicated_genomes_prefixed"
OUTFILE="dereplicated_mags.fna"

mkdir -p "$OUTDIR"

> "$OUTFILE"

for MAG in $INDIR/*$EXT; do
    BASE=$(basename "$MAG" "$EXT")
    SAMPLE=$(echo "$BASE" | awk -F '__|_bin' '{print $1}')
    MODIFIED_MAG="$OUTDIR/${BASE}$EXT"
        awk -v PREFIX="$SAMPLE" \
        '/^>/{print ">" PREFIX "__" substr($0, 2); next} \
        {print}' "$MAG" > "$MODIFIED_MAG"
    cat "$MODIFIED_MAG" >> "$OUTFILE"
done
```

### *3.11 - Assigning Taxonomy to Viruses*

Viruses are assigned taxonomy based on genome similarity (99). To obtain taxonomic assignments for viruses from metagenomes, the tool vConTACT2 (32) can be used. vConTACT2 clusters input viral genomes with reference viral genomes that have taxonomic assignments by building gene-sharing networks (32). The tool used to identify viral scaffolds earlier, geNomad, provides its own marker-based taxonomic classifications in its output (28). These may be used instead of vConTACT2, but they were made prior to vMAG binning and may not be complete. Also, a unique feature of vConTACT2 is its generation of viral clusters (VCs). The VCs generated by vConTACT2 roughly represent virus genera or sub-families (32). VCs are taxonomically and functionally useful, as the viruses within the same VC are closely related in evolution. Even if a set of viral genomes in a VC was not able to be assigned to a described virus taxon, VC-level analyses can be useful when virus species-level analyses are too high in resolution. vConTACT2 does have a notable drawback, it doesn't scale so well with large (> 100,000) viral datasets. However, with patience, vConTACT2 can successfully be run on large datasets. vConTACT3 is also in development, which is reported to be more accurate and scalable than vConTACT2.

We will form VCs and attempt to obtain taxonomic assignments for the dereplicated, species-level vMAGs with vConTACT2. vConTACT2 takes `.faa` protein sequences as input. Genes on viral scaffolds were predicted and translated earlier by geNomad, but the geNomad `.faa` outputs do not contain the reformatted headers with assembly IDs and vMAG names. They also include many proteins from viral scaffolds that were filtered during binning and dereplication. For these reasons, genes will be predicted and translated on the file with dereplicated vMAGs, `virus_genomes_dereplicated.fna`, generated above in Section 3.10. The tool prodigal-gv (28) will be used to annotate the vMAGs in this file. Prodigal-gv is based off of the robust and accurate gene prediction tool, Prodigal (100), but it is optimized for viruses (particularly 'giant viruses' and viruses with stop codon reassignments, but it works well for all kinds of viruses of microbes). The following command is used to run prodigal-gv on the file `virus_genomes_derepliated.fna`:

```
prodigal-gv \
    -i virus_genomes_dereplicated.fna \
    -a virus_genomes_dereplicated.faa \
    -d virus_genomes_dereplicated.ffn \
    -p meta
```

The predicted genes are written to `virus_genomes_dereplicated.ffn` and are translated in the file `virus_genomes_dereplicated.faa`. The flag `-p meta` is necessary to run prodigal-gv in metagenome mode.

Since proteins encoded on multiple scaffolds can belong to a single vMAG, a gene-to-genome table must be made for vConTACT2. This is a three-column table with gene IDs, genome names, and optional keywords in the third. Using the dRep auxiliary script `parse_stb.py` and the python script `make_g2g.py` shown below, this file can be created:

```
parse_stb.py \
    --reverse \
    --fasta dereplicated_viruses/dereplicated_genomes/*.fasta \
    --output virus_genomes_dereplicated
```

```
# make_g2g.py

# Define file paths
faa_file_path = 'virus_genomes_dereplicated.faa'
stb_file_path = 'virus_genomes_dereplicated.stb'
output_file_path = 'virus_genomes_dereplicated.gene_to_genome.csv'

# Load contig to genome mapping from STB file
contig_to_genome = {}
with open(stb_file_path, 'r') as stb_file:
    for line in stb_file:
        contig, genome = line.strip().split()
        genome = genome.rsplit('.', 1)[0]  # Remove the .fasta extension
        contig_to_genome[contig] = genome

# Process FAA file and write gene to genome mapping
with open(faa_file_path, 'r') as faa_file, open(output_file_path, 'w') as out
put_file:
    output_file.write('protein_id,contig_id,keywords\n')
    for line in faa_file:
        if line.startswith('>'):
            # Extract gene name
            gene_name = line.split()[0][1:]
            # Extract contig name from gene name
            contig_name = '_'.join(gene_name.split('_')[:-1])
            # Map contig to genome name
            genome_name = contig_to_genome.get(contig_name, "Unknown")
            output_file.write(f'{gene_name},{genome_name},\n')

print("Done", output_file_path)
```

Finally, vConTACT2 can be executed using the protein sequences in `virus_genomes_dereplicated.faa` and the newly generated gene-to-genome file `virus_genomes_dereplicated.gene_to_genome.csv`:

```
vcontact2 \
    --raw-proteins virus_genomes_dereplicated.faa \
    --proteins-fp virus_genomes_dereplicated.gene_to_genome.csv \
    --db ProkaryoticViralRefSeq207-Merged \
    --output-dir vcontact \
    --threads 16
```

When vContACT2 is finished running, there will be two main output files of interest in the folder `vcontact`. The file `genome_by_genome_overview.csv` is a table with all reference and user-provided viral genomes. The taxonomy for every genome will be given from the Kingdom to the Genus level, if available, as well as information for the VC in which the genomes were clustered. The file `c1.ntw` is a network table, wherein the first two columns are viral genome names, and the third column is an edge weight that connects them. This weight is associated with the clustering results. Using the `genome_by_genome_overview.csv` output as a node table and the `c1.ntw` file as an edge table, a gene-sharing network of viral genome relatedness can be visualized in a desktop program such as Cytoscape (101).

### *3.12 - Predicting the Microbial Hosts of Viruses*

Identifying host-virus interactions in microbial communities can offer significant insights on the impact of viruses on microbial community dynamics, the roles of viruses in the biogeochemistry of ecosystems (4, 102), and the flow of genetic information in a community through horizontal gene transfer (10). In studies with only metagenomic sequence data available, it is difficult to predict the host of viruses, especially highly novel ones. Approaches like single-cell sequencing and Hi-C metagenomics can be used to supplement metagenomic data for predicting the hosts of uncultivated viruses (103, 104), but predicting the microbial hosts of viruses from metagenomic data alone represents a fundamental challenge in microbial and viral ecology. There have been a variety of strategies used to predict the hosts of viruses in metagenomes. Host-based tools detect relationships through sequence alignments between phage and host genomes, which can indicate prophages in host genomes (105) or match phage genomes to host CRISPR spacers (106). Other, alignment-free host-based tools compare k-mer frequencies between viruses and microbial genomes, which may indicant an adaptation in viral genomes to their host genomes (107, 108). Phage-based tools, on the other hand, leverage databases of reference viruses with known hosts, offering predictions based on the similarity to the known viruses (109, 110).

The tool iPHoP (integrated Phage-Host Prediction) (33) integrates both phage-based and host-based predictive approaches from multiple tools to gain predictions down to the host genus level for both known and novel phages. It comes with an expansive database of prokaryotic genomes from GTDB (79) to use as potential hosts. If the user has their own set of MAGs, perhaps from the same samples as the query viruses, the MAGs can be combined with iPHoP's database to potentially obtain precise host predictions to the MAGs of interest. This is a powerful feature of iPHoP. To build this custom database, iPHoP requires the MAGs to be placed into a phylogenomic tree with the GTDB genomes using the GTDB-tk *de novo* workflow. This is different than the GTDB-tk classify workflow used to assign taxonomy to the MAGs, above.

Two trees will need to be built with the medium- and high-quality, dereplicated, *Tara* MAGs, one for archaea and one for bacteria. The command `gtdbtk de_novo_wf` will be used to create these in a new folder, `host_predictions`:

```
mkdir host_predictions
```

```
gtdbtk de_novo_wf \
    --archaea \
    --cpus 16 \
    --genome_dir dereplicated_mags/dereplicated_genomes \
    --extension fa \
    --out_dir host_predictions/GTDB-tk \
    --outgroup_taxon p__Nanoarchaeota
```

Here, the flag `–archaea` is used to build a tree with GTDB archaeal sequences, `–genome_dir` provides the path to the dereplicated MAGs, `–extension fa` indicates that the MAG filenames end with the extension `.fa`, `–out_dir host_predictions/GTDB-tk` will put the GTDB-tk results in the folder GTDB-tk beneath the parent folder `host_predictions`, and importantly, `–outgroup_taxon p__Nanoarchaeota` sets the outgroup taxon of the tree to genomes in the phylum Nanoarchaeota. When running `gtdbtk de_novo_wf`, it is important to pick an outgroup phylum that is not represented in your dataset of input MAGs. In several cases, Nanoarchaeota may be a safe choice, but this should not be blindly trusted. If one is studying an environment where it was likely to recover Nanoarchaeota MAGs, a different phylum must be chosen. To be certain of which phyla are present, the previous taxonomy results generated in Section 3.7 can be used, or a simple BLAST of 16S sequences and/or other marker genes may be helpful given their high conservation at the phylum level.

For the bacterial tree, the same command can be executed with some modifications:

```
gtdbtk de_novo_wf \
    --bacteria \
    --cpus 16 \
    --genome_dir dereplicated_mags/dereplicated_genomes \
    --extension fa \
    --out_dir host_predictions/GTDB-tk \
    --outgroup_taxon p__Campylobacterota
```

The results will be deposited in the same folder as the archaeal tree. Here, the outgroup phylum Campylobacterota was chosen since it was not found in the GTDB taxonomy results from Section 3.7.

Once the two GTDB *de novo* trees have been built with the dereplicated MAGs, they can be added to the existing iPHoP database with the `iphop add_to_db` command:

```
iphop add_to_db \
    --fna_dir dereplicated_mags/dereplicated_genomes \
    --gtdb_dir host_predictions/GTDB-tk \
    --out_dir host_predictions/iphop_db_with_custom_mags \
    --db_dir /path/to/default/iphop/database
```

The dereplicated MAGs are given to iPHoP with the flag `–fna_dir dereplicated_mags/dereplicated_genomes`, the GTDB-tk results are given with the flag `–gtdb_dir host_predictions/GTDB-tk`, the output custom database location is specified with `–out_dir host_predictions/iphop_db_with_custom_mags`, and the existing default iPHoP database is given with the `-db_dir` flag, which will have to be set to the path of the user's local copy of the iPHoP database.

Since iPHoP expects one virus per scaffold when predicting hosts, the dereplicated vMAGs will have to be linked together using the vRhyme auxiliary script `link_bin_sequences.py`, just as the non-dereplicated vMAGs were when running CheckV in Section 3.9:

```
link_bin_sequences.py \
    -i dereplicated_viruses/dereplicated_genomes \
    -o dereplicated_viruses/dereplicated_genomes_linked

cat dereplicated_viruses/dereplicated_genomes_linked/*.fasta \
    > virus_genomes_dereplicated_linked.fna
```

Here, the `link_bin_sequences.py` script will join together the scaffolds in each dereplicated vMAG, then write each joined vMAG file to a new folder `dereplicated_genomes_linked` within the `dereplicated_viruses` folder. The `cat` command will combine all of the N-linked dereplicated vMAGs into a single file `virus_genomes_dereplicated_linked.fna`, which contains one 'scaffold' per vMAG. Now, to predict the microbial hosts of the dereplicated vMAGs, this file can be given to the `iphop predict` command:

```
iphop predict \
    --fa_file virus_genomes_dereplicated_linked.fna \
    --out_dir host_predictions/iphop_output \
    --db_dir host_predictions/iphop_db_with_custom_mags \
    --no_qc \
    --num_threads 16
```

The input vMAG file is specified with the `-fa_file` flag, the custom database with GTDB genomes and the dereplicated *Tara* MAGs is given with the `-db_dir` flag, and the output folder with predictions `iphop_output` will be created beneath the `host_predictions` folder using the `-out_dir` flag. Importantly, the flag `-no_qc` is included here. By default, iPHoP excludes input viral genomes with many Ns. Since many of the input vMAGs will have long stretches of Ns from linking scaffolds together, this QC step can be skipped with the `-no_qc` option. This flag is not recommended when running `iphop predict` on single-scaffold viral genomes.

iPHoP will produce two main output files, `Host_prediction_to_genus_m90.csv` and `Host_prediction_to_genome_m90.csv`. The former has the host predictions to the host genus level, integrated from the host-based and phage-based prediction tools. The latter has host predictions at the host genome level, integrated from the host-based tools only. There will (likely) be at most one host genus prediction for a given input vMAG in `Host_prediction_to_genus_m90.csv`, but there can be multiple reference or user-provided genomes that are assigned to that genus in `Host_prediction_to_genome_m90.csv`.

To get a simple count of how many vMAG to host genus pairs were predicted, the number of lines in `Host_prediction_to_genus_m90.csv` can be counted with `wc -l`:

```
wc -l host_predictions/iphop_output/Host_prediction_to_genus_m90.csv

3350
```

To get a count of how many vMAG to host genome pairs were predicted from the *Tara* MAGs, the number of occurrences of a regex pattern that follows the MAG naming scheme can be counted using `grep -cP`:

```
grep -cP "ERR\d\d\d\d\d\d__bin_" host_predictions/iphop_output/Host_predictio
n_to_genomes_m90.csv

56
```

Out of the 156,671 dereplicated vMAGs, host predictions were only obtained for 3,350 of them. A fraction this small is quite typical for environmental MAGs and vMAGs, yet 3,350 host predictions is a substantial number of identified interactions. Although the host genera for most of the vMAGs here are unknown, the identified host-virus pairs are very valuable when leveraged in downstream analyses. The output from iPHoP `Host_prediction_to_genus_m90.csv` can be converted into a network diagram and visualized wth a program like Cytoscape (101) to qualitatively assess host-virus relationships across the *Tara* samples (Figure 5).

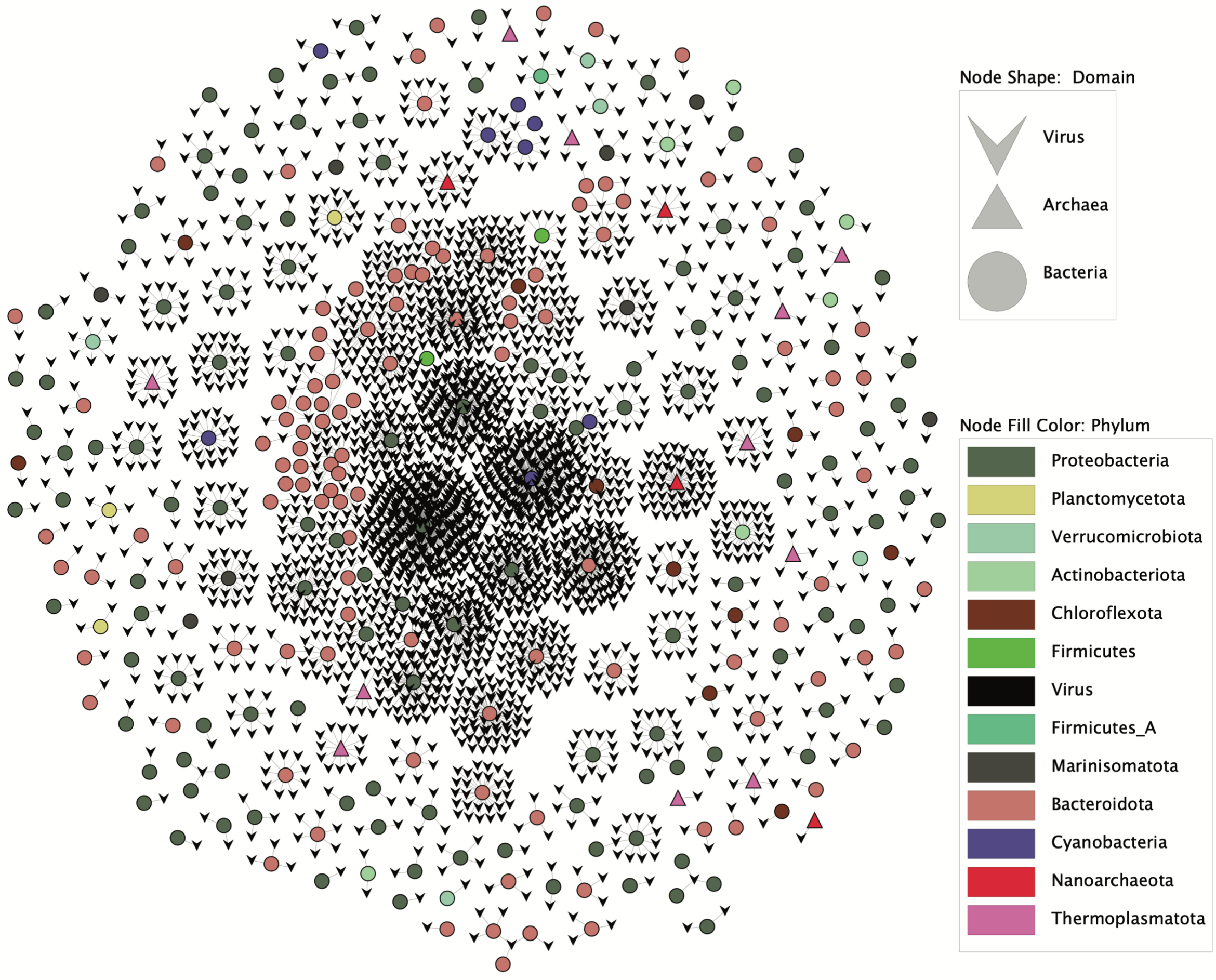

**Figure 5: iPHoP Host Predictions for *Tara* vMAGs.** Host predictions to individual genera are shown, as they appear in the iPHoP output Host_prediciton_to_genus_m90.csv. Host phyla that made up < 1% of all predictions were removed. Network built and visualized with Cytoscape (101).

## 4 - Future Directions for analyses of Microbial and Viral Diversity, Ecology, and Evolution

Section 3 started with a tremendous amount of bulk, unfiltered, unsorted, unassembled raw DNA sequence data that. By itself, it had very little biological meaning, but it had incredible value since it was able to be refined and distilled into very meaningful data as the sections progressed. Now that microbial and viral genomes were reconstructed, filtered, clustered, classified, and associated with each other, we hope that the researcher is equipped with enough data to test their hypotheses. Below, we list a few additional approaches that may be worth pursuing depending on what the researcher is interested in. However, we note that there are virtually endless opportunities that large, multi-sample metagenomic datasets present, so these approaches are only a small slice of the directions one may take with their data.

### *4.1 - Microbial and viral species abundance.*

For microbial communities, alpha- and beta-diversity is typically estimated using 16S amplicon sequencing and inference workflows (111). While marker-based sequencing is the best way to estimate macro diversity of microbes, it is not possible with viruses since they lack universal marker genes. In metagenomics and viromics, species abundance can be estimated by mapping sequences reads to dereplicated genomes. This can be done for both microbes and viruses (4, 112, 113). Although filtered DNA sequence reads were already mapped to their respective assemblies in Section 3.5, those mapping data can only be used to analyze the species within each sample, separately. By creating a mapping index from the MAGs/vMAGs dereplicated across all assemblies, all sequence reads can be mapped to the same index. This will allow the assessment of the abundance and distribution of all microbial/viral species in every sample. Mapping reads to the same, multi-sample genome index can also reveal the presence of species that were not initially detected in a sample due to poor assembly or filtering. Multi-sample read mapping is therefore a critical step to measure microbial and viral species abundance patterns both within and across communities.

### *4.2 - Comparative genomics and microdiversity.*

Species abundance patterns for microbes and viruses can reveal a lot about the diversity of microbiomes at the community level. To bioinformatically analyze the genetic diversity of microbes and viruses within populations or at the genome level (microdiversity), additional approaches need to be employed. Just as they were useful for estimating species abundance patterns, dereplicated genomes and their read-mapping data can be applied here as well. Single-nucleotide variants, insertions, and deletions can be detected via read alignments in BAM files. Some useful tools that employ a pipeline to analyze microdiversity and calculate evolutionary statistics are inStrain (67) and MetaPop (97). They are specifically designed to handle large, multi-sample metagenomic datasets and report many useful statistics on the genetic variation of individual genes and genomes. With these tools, or similar tools, the researcher may see how the genomes of specific microbial or viral taxa in their data vary across their samples. If the researcher has metagenomic data across multiple time points, these tools can be very powerful to test hypotheses on how their microbes and viruses of interest are evolving.

### *4.3 - Functional Analyses of Microbial and Viral Biogeochemistry.*

Both microbes and viruses have significant impacts on their surrounding environments (85, 114, 115). Although non-computational approaches are well suited to test hypotheses on the biogeochemistry of microbiomes, there are several bioinformatic approaches that can assess the contributions of microbes and viruses to the chemistry of their environment. These bioinformatic approaches are particularly useful when studying the biogeochemistry of microbiomes with uncultivated members. METABOLIC (116) is a powerful tool that can take (meta)genomes and predict the metabolic pathways of microbes at the community- and population-level. It can also predict the contribution of individual microbial taxa to specific metabolic pathways in a community. Also, DRAM and DRAM-v (117) is a pipeline that annotates metabolic genes in both microbes and viruses in metagenomic data. Like METABOLIC, it can assess the presence of key metabolic genes and infer the completeness of specific metabolic pathways in taxa of interest. Leveraging tools like METABOLIC or DRAM/DRAM-v can be very useful in microbiome studies with biogeochemically or metabolically related hypotheses.

## 5 - References